%% file: fado.tex
\documentclass[sigconf,review=false,screen=false,authorversion=true,anonymous=false]{acmart}
\usepackage{multirow}
\usepackage[linesnumbered,ruled,vlined]{algorithm2e}

\SetCommentSty{mycommfont}

\makeatletter
\newcommand*\myalgsize{%
	\@setfontsize\myalgsize{7.5}{9.0}%
}
\makeatother

\SetKwInput{KwInput}{Input}                
\SetKwInput{KwOutput}{Output}              
\usepackage[noend]{algorithmic}
\usepackage{amsmath}
\usepackage{mathtools}
\usepackage{hyperref}
\usepackage{balance}
\DeclareMathOperator*{\argmax}{argmax}

\DeclareMathOperator*{\sgn}{sgn}
\DeclareMathOperator*{\minimize}{minimize}

\AtBeginDocument{%
	}

\copyrightyear{2023}
\acmYear{2023}
\setcopyright{acmcopyright}
\acmConference[FPGA '23]{Proceedings of the 2023 ACM/SIGDA International Symposium on Field Programmable Gate Arrays}{February 12--14, 2023}{Monterey, CA, USA}
\acmBooktitle{Proceedings of the 2023 ACM/SIGDA International Symposium on Field Programmable Gate Arrays (FPGA '23), February 12--14, 2023, Monterey, CA, USA}
\acmPrice{15.00}
\acmDOI{10.1145/3543622.3573188}
\acmISBN{978-1-4503-9417-8/23/02}
\begin{document}
	
	\title{FADO: \underline{F}loorplan-\underline{A}ware \underline{D}irective \underline{O}ptimization\\
		for High-Level Synthesis Designs on Multi-Die FPGAs}
	
	\author{Linfeng Du}
	\email{linfeng.du@connect.ust.hk}
	\orcid{0000-0002-3007-4890}
	\affiliation{
		\institution{Hong Kong University of Science and Technology}
		\city{Kowloon}
		\country{Hong Kong}
	}
	
	\author{Tingyuan Liang}
	\email{tliang@connect.ust.hk}
	\orcid{0000-0002-0390-2320}
	\affiliation{
		\institution{Hong Kong University of Science and Technology}
		\city{Kowloon}
		\country{Hong Kong}
	}
	
	\author{Sharad Sinha}
	\email{sharad@iitgoa.ac.in}
	\orcid{0000-0002-4532-2017}
	\affiliation{
		\institution{Indian Institute of Technology Goa}
		\city{Goa}
		\country{India}
	}
	
	\author{Zhiyao Xie}
	\email{eezhiyao@ust.hk}
	\orcid{0000-0002-4442-592X}
	\affiliation{
		\institution{Hong Kong University of Science and Technology}
		\city{Kowloon}
		\country{Hong Kong}
	}
	
	\author{Wei Zhang}
	\email{eeweiz@ust.hk}
	\orcid{0000-0002-7622-6714}
	\affiliation{
		\institution{Hong Kong University of Science and Technology}
		\city{Kowloon}
		\country{Hong Kong}
	}
	
	\renewcommand{\shortauthors}{Linfeng Du, Tingyuan Liang, Sharad Sinha, Zhiyao Xie, \& Wei Zhang}
	
	\input{sec-abstract}

	\ccsdesc[500]{Hardware~High-level and register-transfer level synthesis}
	\ccsdesc[500]{Hardware~Partitioning and floorplanning}
	
	\keywords{High-Level Synthesis, Design Space Exploration, Multi-Die FPGA, Directive Optimization, Floorplanning}

	\setlength{\textfloatsep}{0pt}
	
	\maketitle
	
	\input{sec-introduction}
	\input{sec-relatedwork}
	\input{sec-motivation}
	\input{sec-formulation}

\input{sec-implementation}
	\input{sec-results}
	\input{sec-conclusion}
	\begin{acks}
		\vspace{-0.3\baselineskip}
		This work is partially supported by the RGC GRF grant 16215319, and ACCESS --- AI Chip Center for Emerging Smart Systems, Hong Kong SAR. We would like to thank all the anonymous reviewers for their valuable comments.
	\end{acks}
	
	\bibliographystyle{ACM-Reference-Format}
	\balance
	\bibliography{ref}

\end{document}

%% file: sec-abstract.tex
\begin{abstract}

Multi-die FPGAs are widely adopted to deploy large-scale hardware accelerators. Two factors impede the performance optimization of high-level synthesis (HLS) designs implemented on multi-die FPGAs. On the one hand, the long net delay due to nets crossing die-boundaries results in an NP-hard problem to properly floorplan and pipeline an application. On the other hand, traditional automated searching flow for HLS directive optimizations targets single-die FPGAs, and hence, it cannot consider the resource constraints on each die and the timing issue incurred by the die-crossings. Further, it leads to an excessively long runtime to legalize the floorplanning of HLS designs generated under each group of configurations during directive optimization due to the large design scale.

To co-optimize the directives and floorplan of HLS designs on multi-die FPGAs, we propose the FADO framework, which formulates the directive-floorplan co-search problem based on the multi-choice multi-dimensional bin-packing and solves it using an iterative optimization flow. For each step of directive optimization, a latency-bottleneck-guided greedy algorithm searches for more efficient directive configurations. For floorplanning, instead of repetitively incurring global floorplanning algorithms, we implement a more efficient incremental floorplan legalization algorithm. It mainly applies the worst-fit strategy from the online bin-packing algorithm to balance the floorplan, together with an offline best-fit-decreasing re-packing step to compact the floorplan, followed by pipelining of the long wires crossing die-boundaries.

Through experiments on a set of HLS designs mixing dataflow and non-dataflow kernels, FADO not only well-automates the co-optimization and finishes within 693X$\sim$4925X shorter runtime, compared with DSE assisted by global floorplanning, but also yields an improvement of 1.16X$\sim$8.78X in overall workflow execution time after implementation on the Xilinx Alveo U250 FPGA.

\end{abstract}

%% file: sec-introduction.tex
\vspace{-0.5\baselineskip}
\section{Introduction}\label{sec-introduction}

Guided by optimization directives, high-level synthesis (HLS) compiles high-level behavioral specifications to register-transfer level (RTL) structures, supporting the ever-growing functional and structural complexity of hardware accelerators. The various directives contribute to large design space to search upon. For example, there are 26 directives in Xilinx Vitis HLS 2020.2~\cite{xilinx2020vitis}, each of which has a set of parameters and can be applied at different levels or structures of the HLS source code. Previous works~\cite{3041219, 5158106, pham2015exploiting, piccolboni2017cosmos, wang2020efficient, sohrabizadeh2021enabling, wu2021ironman, ye2021scalehls, liang2019hi, zhao2017comba} mainly use automated design space exploration (DSE) algorithms to search for the Pareto-optimal directive configurations, targeting the lowest latency (execution time in clock cycle) under a specific resource constraint.

To deploy large-scale HLS designs on FPGAs, with consideration of chip yield in fabrication, larger FPGAs with multiple dies emerge based on 2.5D/3D integration techniques. However, the concomitant long net delay due to nets crossing die-boundaries harms the timing quality of the implemented designs. One of the multi-die packaging technologies is the Stacked Silicon Interconnect (SSI) from Xilinx~\cite{saban2011xilinx}, where a silicon interposer integrates multiple dies, also called super logic regions (SLRs). \cite{chaware2012assembly} states that the super long lines (SLLs) between dies cause $\sim$1 ns delay, while \cite{nasiri2015multiple} states that a typical medium-length routing wire within a single die has 4X$\sim$8X shorter delay under the same manufacturing process. Further, \cite{guo2021autobridge} shows that for HLS dataflow designs, a handshake-based model for task-level parallelism, its floorplanning and pipelining can optimize the maximum achievable frequency (Fmax) on multi-die Alveo FPGAs to at most $\sim$300 MHz, because of the delay of SLLs and long routes detoured by specialized IP blocks close to the I/O banks.

To mitigate the delay penalty on multi-die FPGAs, Xilinx proposes using the floorplanning method~\cite{ug906} to keep critical timing paths on a single SLR. However, the fine-grained gate/cell-level floorplanning is very time-consuming. In comparison, \cite{guo2021autobridge} requires that no function in the HLS dataflow region should spread over multiple SLRs and proposes a coarse-grained method to floorplan HLS functions at the SLR level to accommodate the large-scale dataflow designs and pipeline the wires crossing die-boundaries. 

Min-cut floorplanning focuses on meeting the separate resource constraints on slots divided by SLR boundaries or I/O banks and the SLL number constraints between SLRs. However, an initial min-cut floorplan would not always support the latency-centric optimization of HLS directives. When a function's directive configuration changes, its resource also changes, and the original floorplan could be illegal because of resource over-utilization. For coordinating floorplanning with directive DSE, a simple combination is to solve the global floorplanning repeatedly, e.g., using mixed-integer linear programming (MILP) solver \cite{guo2021autobridge}, whenever there's a new directive applied. This incurs an extended runtime of the DSE flow. Thus, we try to replace the global MILP solution with an incremental legalization algorithm to facilitate a highly-efficient integration of iterative directive search and floorplanning.

\begin{figure}[tbp]
\vspace{-0.8\baselineskip}
    \includegraphics[width=0.98\linewidth]{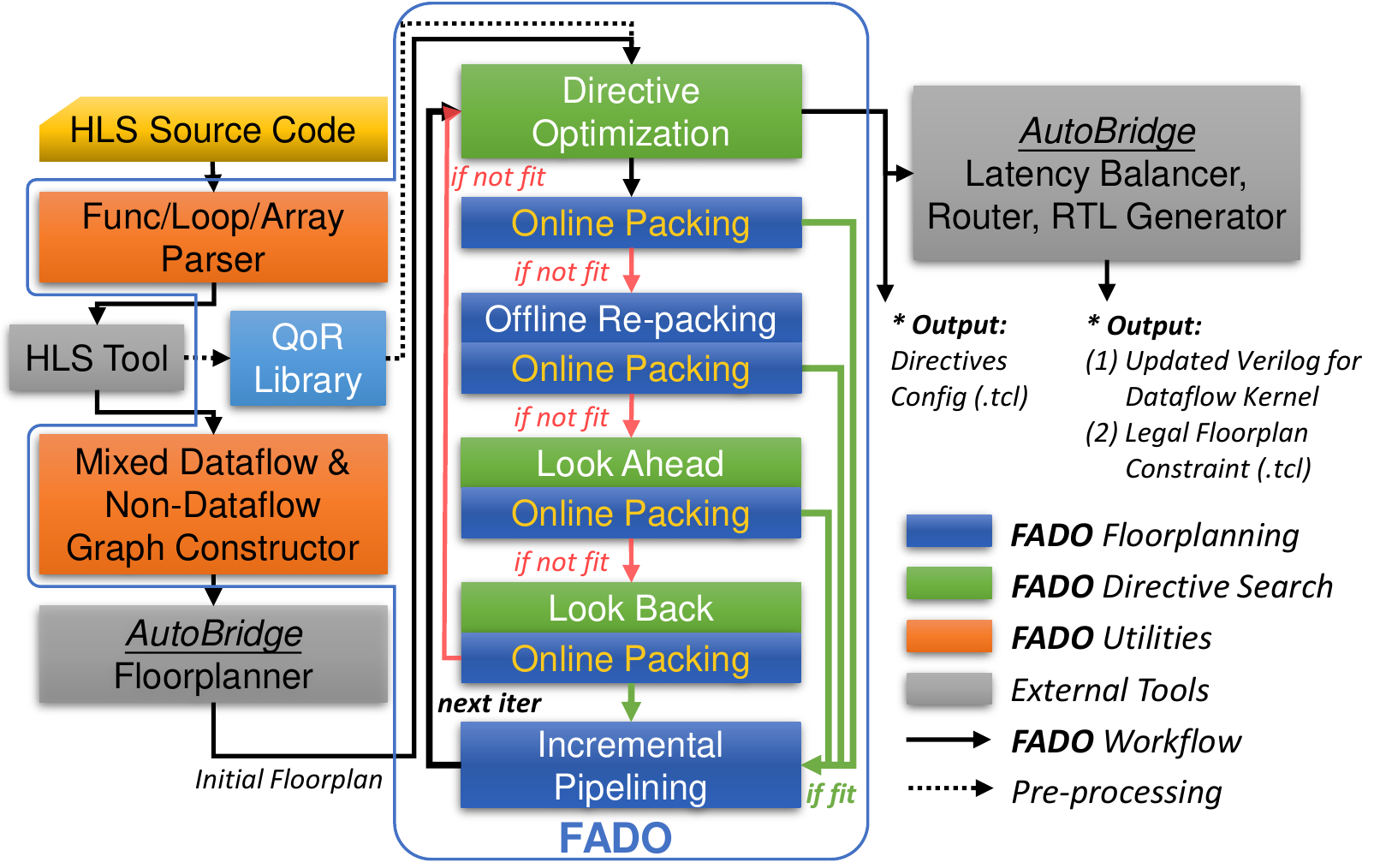}
    \vspace*{-1.1\baselineskip}
    \caption{ Overview of Our FADO Framework. }
    \label{fig:1-framework}
\end{figure}

In this paper, we solve this challenge with a new framework named FADO~\footnote{The FADO framework for directive-floorplan co-optimization is fully open-sourced in \url{https://github.com/RipperJ/FADO}.}, as shown in Fig.~\ref{fig:1-framework}. It is the first work to co-optimize HLS directives and floorplanning on multi-die FPGAs, thus benefiting both latency and timing of HLS designs. We first formulate this complex co-optimization problem based on multi-choice multi-dimensional bin-packing (MMBP)~\cite{patt2010vector}, then develop an extremely efficient iterative solution. In each iteration, we apply a latency-bottleneck-guided greedy algorithm to search for more efficient directive configurations, followed by incremental floorplan legalization. Such legalization applies both worst-fit (WF) online bin-packing algorithm and the best-fit decreasing (BFD) offline algorithm~\cite{man1996approximation}. The WF algorithm balances the resource utilization among slots on FPGA to avoid congestion, while the BFD algorithm breaks the balance with minimum cost to enable the floorplanning of overlarge HLS functions. At the end of each iteration, FADO incrementally adds/updates/removes the pipeline logic along the long wires crossing die-boundaries. This incremental floorplan update is much faster than global algorithms in previous works~\cite{guo2021autobridge}.

In summary, FADO improves the overall performance of designs deployed on multi-die FPGAs with co-optimization and achieves high speed with its customized iterative solution. In our experiment, FADO can fully utilize resources on FPGA under all constraints. Also, FADO proves to scale well to large accelerators with both dataflow and non-dataflow kernels. 

Our contributions in FADO are as summarized below:
\begin{itemize}
    \item To the best of our knowledge, FADO is the first solution for co-optimization of HLS directives and floorplanning on multi-die FPGAs. It improves both latency and timing of complex implemented designs within a very short runtime.
    \item We propose the first precise mathematical formulation of this directive-floorplan co-optimization problem on multi-die FPGAs in FADO. Its solution is based on a well-customized iterative algorithm, which finishes in seconds, achieving orders-of-magnitude speedup over global algorithms in prior works, while producing even better final design quality.
    \item Compared with the directive DSE with the global MILP floorplanning and pipelining~\cite{guo2021autobridge}, FADO achieves 693X$\sim$4925X speedup with an even higher and near-optimal final design performance on FPGA for all six tested designs. The performance measured with overall workload execution time on each design improves 1.16X$\sim$8.78X.
    \item Compared with \cite{guo2021autobridge}, our FADO framework can automatically handle not only dataflow benchmarks but also large-scale applications mixing dataflow and non-dataflow kernels.
\end{itemize}

\vspace{-0.4\baselineskip}

%% file: sec-relatedwork.tex
\section{Related Work}\label{sec-relatedwork}
\vspace{-0.1\baselineskip}

\begin{table}[!tbp]
\vspace{-0.8\baselineskip}
    \caption{Comparisons between FADO and Previous Work \vspace{-1\baselineskip}}
    \setlength\tabcolsep{3pt}
    \resizebox{0.91\linewidth}{!}{
    \begin{tabular}{ c c c c }
    \toprule
     & Directive & Multi-die  & Floorplan-aware \\
     & Search &  Floorplanning    & Directive DSE (\textbf{FADO})   \\
    \midrule
    \multirow{2}{*}{QoR} & latency, & timing        & latency, resource, \\
                         & resource & (frequency)   & timing \\ 
    \hline
    Design  & 1. directives & 2. SLR-level & \multirow{2}{*}{1 \& 2} \\
    Space   & \& parameters    &  func location    &          \\
    \hline
    DSE          & syn: slow;        & slow (SA, MILP,       & fast (incremental \\
    Efficiency   & model: fast      & bi-partition, ...)    & floorplanning)\\    
    \hline
    Type of      & dataflow or      & \multirow{2}{*}{dataflow} & mixed dataflow \\
    Benchmark   & non-dataflow     &                           & \& non-dataflow \\  
    \bottomrule
    \end{tabular}}
    \label{tab:2-related-work}
\vspace{0.3\baselineskip}
\end{table}

As shown in Table.~\ref{tab:2-related-work}, FADO performs co-optimization considering latency, resource, and timing during floorplanning, while previous commonly used flows, such as directive search and multi-die floorplanning, only target one or two optimization objectives. To achieve multi-objective optimization, the design space is enormous, defined by the Cartesian product of directives, their respective parameters, and SLR-level function locations. Previous works may take a long time to traverse such a large design space. However, with our effective incremental floorplanning, FADO achieves orders-of-magnitude speedup in the search time. Moreover, existing floorplanning works for HLS designs~\cite{guo2021autobridge, alonso2021elastic} are dedicated to dataflow applications. FADO is able to automatically solve the floorplanning for non-dataflow functions by adding additional constraints.

\newpage

\noindent\textbf{HLS Directive Optimization} has been researched thoroughly. The featured challenges are listed below.

\begin{itemize}
	\item Directives and their parameters contribute to an enormous design space~\cite{sohrabizadeh2022autodse}. Accordingly, we apply a latency-bottleneck-guided algorithm to minimize the overall latency and speed up the DSE effectively.
	\item Directive configurations have non-monotonic effects~\cite{sohrabizadeh2022autodse} on QoRs, which is also explained by~\cite{yu2018s2fa} as inter-dependency among directives and structures in the source code. To avoid getting trapped in local optima on the non-monotonic design space, we design the look-ahead/back sampling methods. 
\end{itemize}


The general techniques for HLS directive DSE includes (1) meta heuristics~\cite{3041219, 5158106, xydis2010efficient}, (2) dedicated heuristics~\cite{pham2015exploiting, piccolboni2017cosmos, yu2018s2fa}, (3) machine learning~\cite{meng2016adaptive, wang2020efficient, zacharopoulos2018machine}, and (4) graph analysis~\cite{sohrabizadeh2021enabling, wu2021ironman, ye2021scalehls, zhao2017comba}. Prior work on directive search mainly optimizes latency under an overall resource constraint of single-die FPGAs. To compare, multi-die FPGAs introduce separate constraints on each slot and between SLRs, and also the vital timing issue because of long wires crossing die-boundaries. Hence, we cannot directly apply the previous DSE algorithms to our co-optimization problem. 



\begin{table}[!t]
\vspace{-0.3\baselineskip}
    \caption{Objectives of Multi-die FPGA Timing Optimizations \vspace{-1\baselineskip}}{
    \resizebox{0.6\linewidth}{!}{
    \begin{tabular}{ c c } 
        \toprule
        Objectives & Previous Works\\
        \midrule
        Total Wirelength & \cite{deng2003physical, mao2016modular, zhang2019frequency} \\
        Signal Delay & \cite{mao2016modular, hahn2014cad, nasiri2015multiple} \\
        Number of Cut Net (SLL) & \cite{guo2021autobridge, hahn2014cad, nasiri2015multiple, ravishankar2018placement} \\
        Routing Congestion & \cite{hahn2014cad, nasiri2015multiple, ravishankar2018placement, alonso2021elastic} \\
        Aspect Ratio & \cite{deng2003physical, ravishankar2018placement} \\
        \bottomrule
    \end{tabular}}
    }
    \label{tab:2-partition-obj}
\end{table}

\noindent\textbf{Multi-die FPGA Timing Optimization} can be classified by their objectives as Table.~\ref{tab:2-partition-obj} shows. \cite{deng2003physical} proposes optimization on total wirelength and aspect ratio of face-to-face-stacked floorplans. \cite{mao2016modular} minimizes total wirelength while reducing total and die-crossing delay. \cite{zhang2019frequency} proposes constructive floorplan optimizations dedicated to PEs of systolic arrays. \cite{hahn2014cad} and \cite{nasiri2015multiple} extend the P\&R tool VPR~\cite{betz1997vpr} to multi-die scenario by adding parameters to the cost function including wire-cut ratio, delay increment, and cut number, while \cite{nasiri2015multiple} also considers the congestion cost. \cite{guo2021autobridge} applies MILP to minimize the number of die-crossing long wires. It runs iterative bi-partitioning rather than N-way partitioning and prefers the most balanced floorplan across all slots divided by die-boundaries and I/O banks. \cite{ravishankar2018placement} implements a partition-driven placer and an aspect-ratio-aware cut scheduling algorithm. \cite{alonso2021elastic} also applies ILP and resource balancing heuristics to partition dataflow accelerators and average congestion among different SLRs. It also discusses partitioning dataflow accelerators among multiple FPGAs through network interfaces. 

In FADO, we mainly compare with \cite{guo2021autobridge} on the efficiency of coarse-grained floorplanning because fine-grained floorplanning in other works is too time-consuming for large-scale designs, not to mention floorplanning repetitively during iterative DSE. We identify that the major frequency improvement in \cite{guo2021autobridge} comes from the insertion of pipelining logic, while min-cut floorplanning mainly performs legalization for logic resources and die-crossings. Thus, we replace the timing-consuming min-cut MILP floorplanning with an incremental legalization algorithm with a partial pipelining update to speed up the DSE without sacrificing floorplanning quality.


\noindent\textbf{Knapsack}~\cite{martello1990knapsack} and \textbf{Bin-Packing Problems}~\cite{martello1990bin} are a series of classic combinatorial optimization problems having a common ground with the directive-floorplan co-search. The basic version is the 0-1 Knapsack problem, where multiple items with different weights and values are to be packed in a knapsack, and a binary choice is made for packing the item or not. \cite{sinha1979multiple, kellerer2004multiple} introduces the multiple-choice Knapsack problem, where the items are classified, and exactly one from each class is chosen to form a solution. The classes here map to the directive configurations for an HLS function in our problem. \cite{kulik2010there} introduces the multiple-dimensional Knapsack problem, where the weight of each item and the capacity of knapsacks are in vectors. This corresponds with the types and amounts of resources on each die of a multi-die FPGA. \cite{akbar2001heuristic, patt2010vector} separately formulates the multi-choice multi-dimensional Knapsack problem (MMKP) and bin-packing problem (MMBP). The optimal solution to this problem can be found using branch-and-bound with linear programming, but the high time complexity does not support a large number of variables and equations. Another approximation is using greedy approaches, generally sorting items based on the values and the weights in a certain order. We thus formulate our problem based on the MMBP and combine online Worst Fit (WF) and offline Best-Fit Decreasing (BFD)~\cite{man1996approximation} bin-packing heuristics to solve it efficiently. Here, in the online algorithm, the decision of packing an item is irreversible, and the next item is only visible after the previous packing is settled. For offline algorithms, the value and weight of all items are visible from the very beginning, and we can sort the items to improve the packing quality. \cite{karp1992line}
\vspace{-0.7\baselineskip}

%% file: sec-motivation.tex
\section{Motivation}\label{sec-motivation}

\begin{figure*}[tbp]
\vspace{-0.7\baselineskip}
    \includegraphics[width=\textwidth]{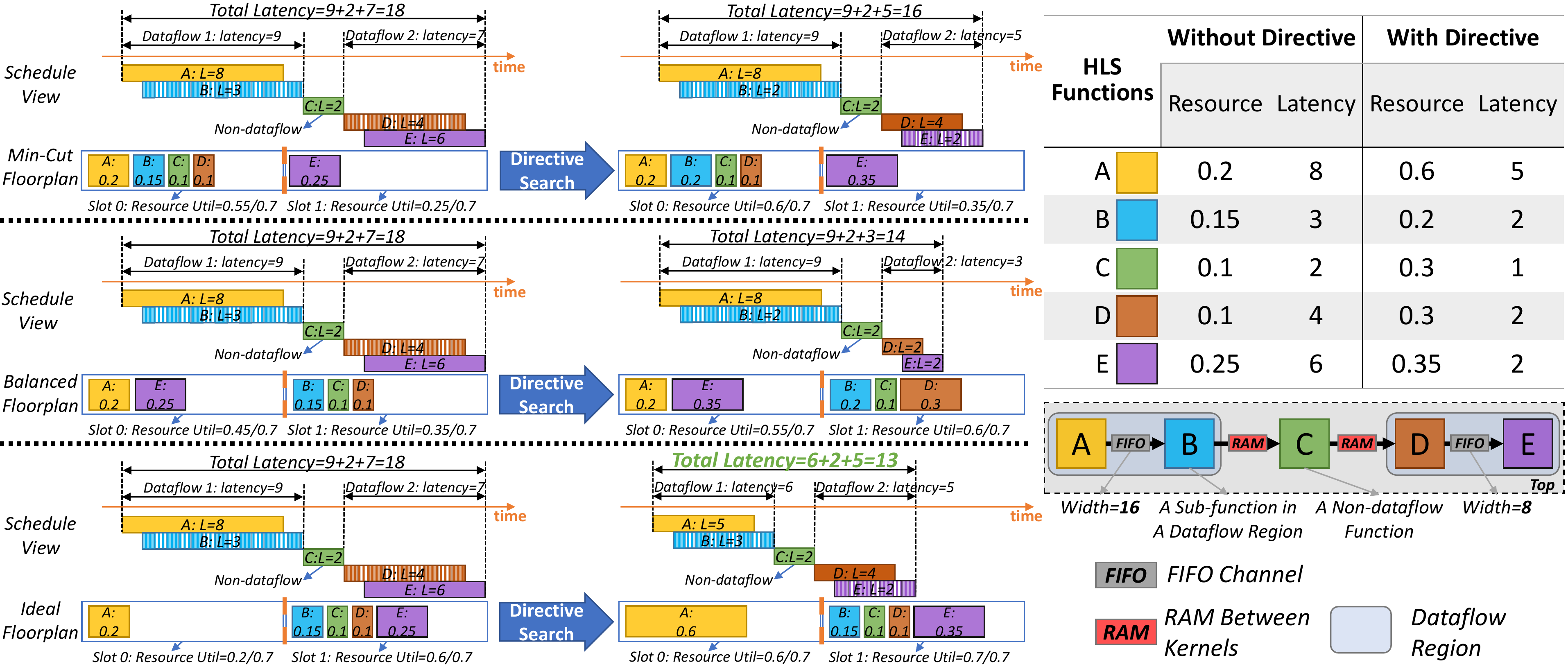}
    \vspace{-1.8\baselineskip}
    \caption{A Toy Example with 2 Dataflow Kernels and 1 Non-dataflow Kernel, with Different Latency (L) and Resource Consumption (R). Three Different Floorplanning Methods and the Corresponding DSE Results Are Compared. \vspace{-1.1\baselineskip}}
    \label{fig:3-motivation}
\end{figure*}

To show the different challenges of the directive-floorplan co-search problem on multi-die FPGAs, we use a toy example with one step of floorplanning followed by directive optimization to explain why general floorplanning algorithms and heuristics won't collaborate with directive search.

Suppose that we have a toy multi-die FPGA with two slots, each having a constraint of $70\%$ of the total available resource, as suggested by \cite{guo2021autobridge, alonso2021elastic}. Although there are several different resources on a modern FPGA, such as Look-up Tables (LUT), Flip-Flops (FF), Digital Signal Processors (DSP), Block-RAMs (BRAM), UltraRAMs~\cite{ultraram} (URAM), etc., in our experiments, we normalize and take the maximum among all resources as the final utilization ratio in evaluation.

Fig.~\ref{fig:3-motivation} shows a design consisting of 2 dataflow and 1 non-dataflow kernel. A/B (or D/E) are functions connected by the FIFO channels in the same dataflow kernel (region), defined as an HLS function with directive "DATAFLOW", achieving task-level parallelism within the function through a handshake-based model. C is a non-dataflow function connected through RAM to other kernels. The channel width between A and B is 16, and that between D and E is 8. 

For latency optimization, different QoRs are caused by various directive configurations in HLS. For example, when applying a smaller initiation interval (II) to the directive \textit{PIPELINE}, or a larger factor to the \textit{UNROLL}, the latency of an HLS design tends to decrease, while the resource utilization is likely to increase. In this example, assume that we only have one directive --- two configurations for each function independently, either with or without that directive. When the directive is applied to a function, its resource consumption increases and latency decreases, as shown in Fig.~\ref{fig:3-motivation}.

For timing optimization, a design's frequency can be ensured at a high level by pipelining as long as the following floorplan conditions are satisfied. The first is having a legal floorplan which meets the resource constraint on every single slot. Second, only FIFO channel connections are allowed to cross the slot boundaries (within a limit on total width not reflected in this toy example) because the handshake interface of FIFO is easy to pipeline, while the complex RAM interface cannot be pipelined. Thus, when functions are connected through RAM, they should be grouped and floorplanned on the same slot, while functions connected through FIFO channels can be partitioned on different slots. 

During directive DSE, we minimize the total latency of the design while ensuring the two floorplan conditions above. Suppose that every function has no directive applied at the beginning of DSE. We find an initial floorplan using a specific algorithm first and then try to improve some functions by applying their respective directive, subject to the resource constraint on each slot.

Fig.~\ref{fig:3-motivation} shows the three floorplanning objectives compared in this toy example. The first minimizes the width of the FIFO channel crossing two slots (min-cut), as used in \cite{guo2021autobridge}. Since all functions cannot be packed in one single slot, the solution to the min-cut problem is 8, which is the width of the channel between D and E. Thus, function E should be assigned to the other slot. In this case, functions B and E can still fit into their slots when applying directives, and the final total latency is improved to 16.

The second floorplan refers to the heuristics of resource balancing~\cite{alonso2021elastic}. Since functions B, C, and D are grouped during floorplanning, as the largest one, they are floorplanned onto the other slot initially. After DSE, directives are applied for functions B, D, and E, and the total latency is improved to 14.

To compare, an ideal floorplan for this design is to partition between A and B, and the best point improves total latency from 18 to 13 clock cycles, which is the minimum achievable latency for this case. If we have an ideal floorplan at the very beginning, it's natural for DSE to reach the optimal latency without any effort to change the floorplan. However, if we start with the other two floorplans, neither the min-cut nor balancing algorithm can further improve the achieved latency. In our FADO framework, the incremental floorplanning algorithm smartly re-packs the functions from either of the two prior floorplans and always reaches the ideal solution finally. To be specific, if FADO starts with the min-cut solution, A is identified as the latency bottleneck and has the top priority to apply the directive. When online packing finds no legal floorplan for A under the min-cut floorplan, the offline re-packing stage groups B, C, D, and E. Thus, the ideal floorplan is found, and the directive for A is successfully applied. It's a similar workflow if we start with the balanced floorplan. In Sec.~\ref{sec-results}, the control experiments on real benchmarks all start with a min-cut floorplan, to fairly compare the effectiveness of FADO with MILP floorplanning assisted DSE.

From this example, we want to show that previous floorplanning techniques could fail to assist directive optimization on multi-die FPGAs. On the contrary, an improper floorplan could prune the high-performance points in a design space. That's why we propose floorplan-aware directive optimization, the FADO framework.


%% file: sec-formulation.tex
\DeclarePairedDelimiter{\ceil}{\lceil}{\rceil}
\DeclarePairedDelimiter{\floor}{\lfloor}{\rfloor}

\vspace{-1.6\baselineskip}
\section{Problem Formulation} \label{sec-formulation}

Based on the multi-choice multi-dimensional bin-packing problem, we formulate the directive-floorplan co-search problem on multi-die FPGAs. To accurately describe the problem and show the complexity compared with floorplanning in \cite{guo2021autobridge}, we also present a MILP formulation below. Note that MILP is only used for a description of the problem. In the implementation (Sec.~\ref{sec-implementation}), we are using approximation heuristics for the objective and each constraint instead of repetitively calling the MILP solver.

\vspace{-1.8\baselineskip}
\subsection{Symbol Definition}\label{sec-formulation-sub-symbols}

Table~\ref{tab:4-symbols} shows the domain and definition of all the variables used in our formulation.

\begin{table}[htbp]
    \vspace*{-0.6\baselineskip}
    \setlength\tabcolsep{1.5pt}
    \caption{Symbols Used in Problem Formulation \vspace*{-0.8\baselineskip}}{
    \resizebox{0.85\linewidth}{!}{
    \begin{tabular}{ c l } 
        \toprule
        Symbols & Definition\\
        \midrule
        \multirow{3}{*}{$L_{ij}$} & The latency of the $j$-th function in the $i$-th kernel  \\
                                  & along the longest path. $i\in \{1,2,..., m+n\},$ \\
                                  & $j \in \{1,2,..., C_{i}\}$. $C_i=1, \forall i \geq m+1$. \\
        \hline
        \multirow{3}{*}{$Z_{ijp}$} & Suppose that function $j$ or kernel $i$ has $Q_{ij}$ directive \\
                                   & choices in total. $Z_{ijp}=1$ when directive choice $p$ is  \\
                                   & applied to function $j$ of kernel $i$. $p\in \{1,2,..., Q_{ij}\}$.\\
        \hline
        \multirow{2}{*}{$x_{ijk}$} & $x_{ijk}=1$ when the sub-function $j$ of kernel $i$ is \\
                                   & floorplanned to slot $k\in\{1, 2, ..., S\}$ among $S$ slots. \\
        \hline
        \multirow{3}{*}{$r_{ijpt}$} & $r_{ijpt}$ represents the consumption of resource type $t$ \\
                                    & of function $j$ in kernel $i$, when directive choice $p$ \\
                                    & is applied. $t\in {BRAM, DSP, FF, LUT, URAM}$.\\
        \hline
        $R_{kt}$                   & The total amount of resource $t$ on slot $k$. \\
        \hline
        \multirow{4}{*}{\shortstack{$e_{i1,j1,i2,j2,R}$\\$e_{i1,j1,i2,j2,F}$}}   & $e_{i1,j1,i2,j2,*} \in \mathbb{N}$. It's positive when there exists a RAM \\
                                   & or FIFO connection from function $j1$ in kernel $i1$, to $j2$ \\
                                   & in kernel $i2$, with a width of $e_{i1,j1,i2,j2,*}$. $i1, i2\in$ \\
                                   & $\{1,2,..., m+n\}, j1\in \{1,2,...,C_{i1}\}, j2\in \{1,2,...,C_{i2}\}$.\\
        \hline
        $W, H$                      & Width/Height of an FPGA (by number of slots). \\
        \hline
        \multirow{4}{*}{${b_{h}}_{(k, ks, kd)}$} & ${b_{h}} \in \{0, 1\}$, it equals $1$ when a source function on slot \\
                                    & $ks$ goes through the die boundary indexed with $k$ \\
                                    & and connects to a destination function on slot $kd$. \\
                                    & $k=(k_x, k_y), k_x \in \{0, 1, ..., W-1\}, k_y \in \{0, 1, ..., H-2\}$. \\
        \hline
        $B_h$                       & Total amount of SLLs on a die boundary. \\
        \bottomrule
    \end{tabular}}
    }
    \label{tab:4-symbols}
\end{table}

\vspace*{-1.8\baselineskip} 

\subsection{MILP Formulation}\label{sec-formulation-sub-equations}

The objective function in our problem minimizes the total latency of an HLS design. 
To show the generality of our problem, assume that we have a large accelerator containing multiple dataflow and non-dataflow kernels connected by RAMs. Minimizing the total latency is equivalent to minimizing the sum of latency of kernels $L_{i}$ along the longest path, as Eq.~\ref{eq:4-formulation-eq1} shows.

\vspace*{-1\baselineskip}
\begin{equation}\label{eq:4-formulation-eq1}
 {\minimize \sum_{i=1}^{N} L_{i}}
\end{equation}
\vspace*{-0.9\baselineskip}

Generally, a dataflow kernel's total latency is very close to the longest sub-function $\max_{j} L_{ij}$ in it, meanwhile relatively much smaller latency comes from the depth of dataflow --- $\max_{j} L_{ij} \gg L_{depth}$. We here approximate the total latency of a dataflow kernel $i$ with the latency of the longest sub-function.

For the longest-latency path with $m$ dataflow kernels and $n$ non-dataflow kernels, the objective function can be re-written as:

\vspace*{-1.2\baselineskip}
\begin{equation}\label{eq:4-formulation-eq2}
    \minimize \sum_{i=1}^{m} \max_{j} L_{ij} + \sum_{i=m+1}^{m+n} L_{i}
\end{equation}
\vspace*{-0.9\baselineskip}

where $i$ iterates on kernels, and $j$ on sub-functions. There's no sub-function in non-dataflow kernels indexed from $m+1$ to $m+n$.

\begin{table}[htbp]
\vspace*{-1.1\baselineskip}
    \centering
    \caption{Directives in the Design Space of FADO \vspace*{-0.9\baselineskip} }{
    \resizebox{\linewidth}{!}{
    \begin{tabular}{ c l } 
        \toprule
        Directives & Parameters\\
        \midrule
        PIPELINE & Initiation Interval (II) (<int>: $\{MinII, ..., \floor{4\times MinII, IterLat}\}$)\\
        \hline
        UNROLL & Factor (<int>: $\{1, 2, 4, ..., LoopBound\}$)\\
        \hline
        \multirow{2}{*}{ARRAY\_PARTITION} & Type (Block/Cyclic/Complete) \\
                                          & Dimension (<int>)\\
        \hline
        BIND\_STORAGE & Implementation (BRAM/URAM)\\
        \bottomrule
    \end{tabular}}
    }
    \label{tab:4-drctvs}
\vspace*{-2\baselineskip} 
\end{table}

\subsubsection{Constraint 1: multi-choice packing problem.}\label{sec-formulation-subsub-1}

During the directive search, we have multiple choices of directives and parameters for HLS functions, loops, and arrays. The directive design space of FADO is shown in Table~\ref{tab:4-drctvs}. For \textit{PIPELINE}, the lower bound of II, $MinII$, is determined by recurrence and resource analysis, and it's also revealed by the HLS report when applying the minimum value possible for target II, i.e., 1. The upper bound considers the iteration latency and four times the $MinII$. For \textit{UNROLL}, the applicable value for its factor ranges from 1 to the loop bound. For \textit{ARRAY\_PARTITION}, we consider the three types of partitioning schemes and the dimension of an array. For \textit{BIND\_STORAGE}, an array is either implemented using BRAM or URAM.

Every time we trigger the HLS, one group of directives and the corresponding QoR are applied for each function (including the loops and arrays within it), described as the constraint in Eq.~\ref{eq:4-formulation-eq3}.

\vspace*{-0.9\baselineskip}
\begin{equation}\label{eq:4-formulation-eq3}
    \sum_{p=1}^{Q_{ij}} Z_{ijp} = 1, Z_{ijp} \in \{0, 1\}\\
\end{equation}
\vspace*{-1.1\baselineskip}

\subsubsection{Constraint 2: multiple bins.}\label{sec-formulation-subsub-2}

During floorplanning, multi-die FPGA is partitioned into several slots by the die boundaries and I/O banks. Eq.~\ref{eq:4-formulation-eq4} guarantees that there's no duplicated or missing floorplan for each HLS function.

\vspace*{-0.9\baselineskip}
\begin{equation}\label{eq:4-formulation-eq4}
    \sum_{k=1}^{S} x_{ijk} = 1, x_{ijk} \in \{0, 1\}\\
\end{equation}
\vspace*{-1.1\baselineskip}

\subsubsection{Constraint 3: multi-dimensional packing problem.}\label{sec-formulation-subsub-3}

In our problem, a single dimension of resource constraints corresponds with one type of resource on the multi-die FPGAs. The following constraint in Eq.~\ref{eq:4-formulation-eq5} assures that functions on each slot with specific directive configurations respectively will not cause resource overflow in any type of resource.

\vspace*{-1\baselineskip}
\begin{equation}\label{eq:4-formulation-eq5}
    \sum_{i=1}^{m+n}\sum_{j=1}^{C_i} x_{ijk}z_{ijp}r_{ijpt} \leq R_{kt} \\
\end{equation}
\vspace*{-1.1\baselineskip}

\subsubsection{Constraint 4: grouping the RAM-connected functions.}\label{sec-formulation-subsub-4}

\begin{figure}[tbp]
\vspace{-0.6\baselineskip}
    \centering
    \includegraphics[width=0.7\linewidth]{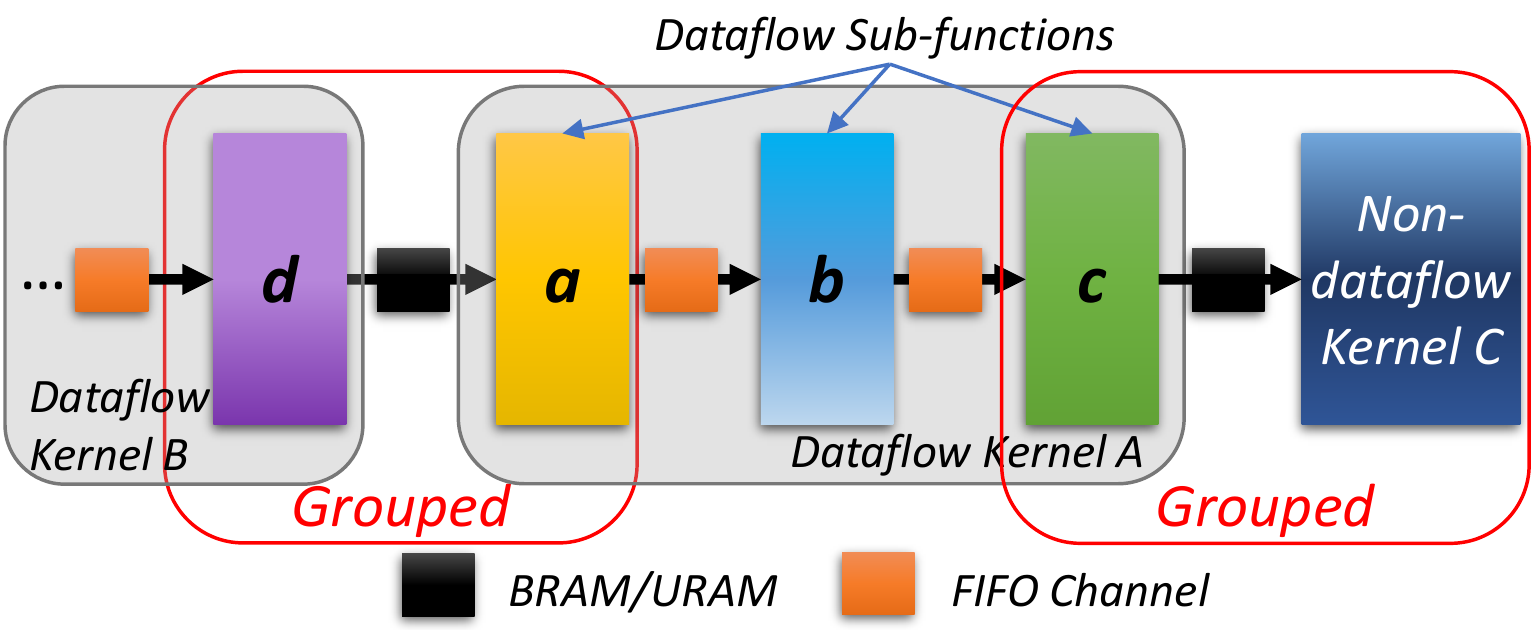}
    \vspace*{-0.8\baselineskip}
    \caption{ Grouped Floorplan for RAM-Interfaced Functions.}
    \label{fig:4-ram}
\end{figure}

Another special type of constraint is introduced by RAM connection between kernels, as shown in Fig.~\ref{fig:4-ram}. Since the interface between RAMs and functions is not the handshake model and is difficult to pipeline, the RAM-connected functions are grouped and assigned to the same slot during floorplanning. As Equation~\ref{eq:4-formulation-eq6} states, $e_{i1,j1,i2,j2,R}x_{i1j1k1}x_{i2j2k2} = 0$ guarantees that either there's no RAM connection between two functions, or they are not separated on two different slots.

\vspace*{-1\baselineskip}
\begin{equation}\label{eq:4-formulation-eq6}
  \begin{aligned}
    & e_{i1,j1,i2,j2,R}x_{i1j1k1}x_{i2j2k2} = 0, (i1 \neq i2 \vee j1 \neq j2) \land k1 \neq k2\\
    & e_{i1,j1,i2,j2,R}\in \mathbb{N_+}, i1 \neq i2 \vee j1 \neq j2\\
  \end{aligned}
\end{equation}
\vspace*{-1.1\baselineskip}


\subsubsection{Constraint 5: Limited number of SLLs.}\label{sec-formulation-subsub-5}

\begin{figure}[tbp]
\vspace{-0.7\baselineskip}
    \setlength{\belowcaptionskip}{-10pt} 
    \centering
    \includegraphics[width=0.7\linewidth]{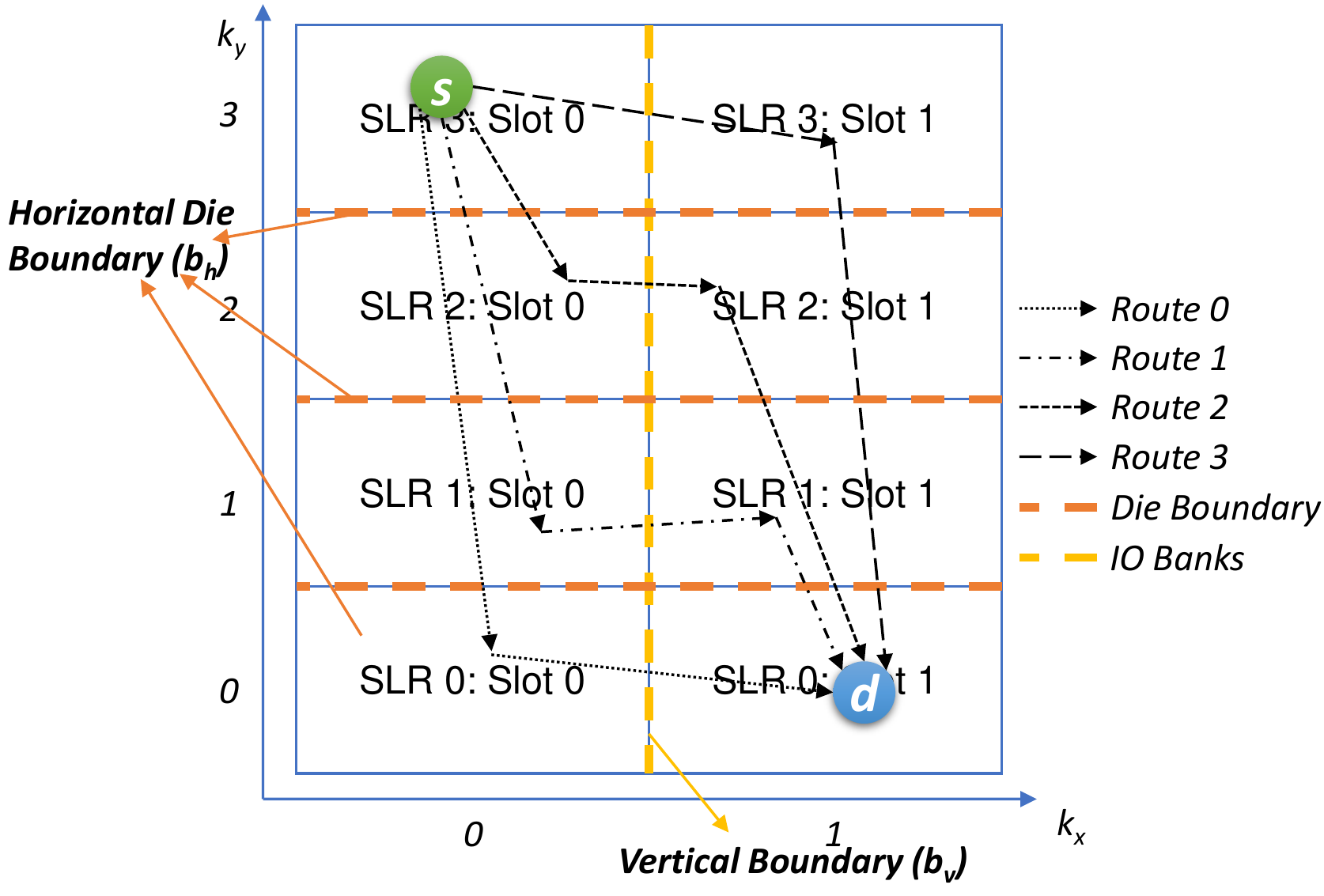}
    \vspace*{-0.8\baselineskip}
    \caption{ An Example about Different Routes between Two Functions on Two Separate Slots of Alveo U250 FPGA. }
    \label{fig:4-SLL}
\vspace{1.2\baselineskip}
\end{figure}

As Fig.~\ref{fig:4-SLL} shows, Alveo U250 FPGA is vertically partitioned into two parts by the I/O banks and horizontally partitioned by die-boundaries into four parts. Suppose we have a source function "s" and a destination function "d" placed on SLR3:Slot0 and SLR0:Slot1, respectively. No matter which route between "s" and "d" is chosen, it crosses three horizontal die boundaries. For each boundary with vertical index $k_y$, the route passes through either the left half or the right half of it. The corresponding constraint is:

\vspace*{-0.9\baselineskip}
\begin{equation}\label{eq:4-formulation-eq7}
    \sum_{k_x=0}^{W-1} x_{i1j1k1}x_{i2j2k2}{b_{h}}_{(k, k1, k2)}\sgn(e_{i1,j1,i2,j2,F}) = 1 \\
\end{equation}
\vspace*{-0.7\baselineskip}

Since the number of SLLs is limited between two dies, we have the formulation in Eq.~\ref{eq:4-formulation-eq8}.

\vspace*{-1.1\baselineskip}
\begin{equation}\label{eq:4-formulation-eq8}
    \sum_{i1,j1,i2,j2,k1,k2} x_{i1j1k1}x_{i2j2k2} {b_{h}}_{(k, k1, k2)}e_{i1,j1,i2,j2,F} \leq \beta B_h,\\
\end{equation}
\vspace*{-0.8\baselineskip}

where $\beta$ is the upper limit of SLL utilization. It is set to $90\%$ in our implementation.

%% file: sec-implementation.tex
\section{Implementation}\label{sec-implementation}

\SetKwFunction{Prune}{Prune}
\SetKwFunction{online}{online\_packing}
\SetKwFunction{offline}{offline\_repacking}
\SetKwFunction{ahead}{look\_ahead}
\SetKwFunction{back}{look\_back}
\SetKwFunction{incrementalFloorplan}{incremental\_Floorplan}
\SetKwFunction{exitcond}{exit\_condition\_check()}


To solve the MILP formulation above by the FADO framework (Fig.~\ref{fig:1-framework}), we describe the objective and multi-choice constraint (Eq.~\ref{eq:4-formulation-eq3}) in the DSE algorithm (Sec.~\ref{directive-search}), and other constraints in the floorplanning algorithms (Sec.~\ref{incr-fp}). The interaction between FADO and external tools is explained in Sec.~\ref{external}.


\vspace*{-0.5\baselineskip}
\subsection{Pre-processing}\label{pre-process}
\vspace*{-0.2\baselineskip}


In a large-scale HLS design, many function blocks could be based on the same template. When running an HLS over the entire design, excessively long synthesis time is wasted on analyzing those functions from the same template repeatedly. Besides, it is hard for existing graph analysis or machine learning methods to accurately predict the QoR of an HLS design with various coding styles, complicated control and data flow, and flexible compilation optimization.

Considering the small size of each function template, it only takes a short period to sample their directive choices and build a function-level QoR library to facilitate the whole workflow. To be specific, we classify the functions by either the template of C++ generics, or custom naming rules, e.g., all functions whose names match the regular expression \textit{r'funcA\_[0-9]\_[0-9]'} will map to \textit{'funcA'} in the QoR library. To apply directives to the C++ template, we follow \cite{xilinx2020template}. As for custom regex, it's straightforward since they have distinct names. Accordingly, we only need to look up for QoRs in the library instead of running HLS flow for an entire design repeatedly.

\vspace*{-0.5\baselineskip}
\subsection{Booting of FADO}\label{booting}
\vspace*{-0.2\baselineskip}

As Fig.~\ref{fig:1-framework} shows, at the very beginning, the input to FADO is an HLS design without any directive of \textit{PIPELINE}, \textit{UNROLL}, \textit{ARRAY\_PARTITION}, and \textit{BIND\_STORAGE}. We parse the source code in "Func/Loop/Array Parser" to generate labels and hierarchy for nested loops, and to identify functions of the same template to build a QoR library $QoR\_lib$ in pre-processing. In the other workflow, the labeled code is synthesized by an external HLS tool. Then, the HLS report is analyzed by a graph constructor, where connections through FIFOs and RAMs are identified. Then, the graph is passed to the min-cut floorplanner of AutoBridge~\cite{guo2021autobridge} to generate an initial legal floorplan, which is then passed on to FADO for directive-floorplan co-optimization.

\begin{figure}[tbp]
\vspace{-0.7\baselineskip}
    \centering
    \includegraphics[width=0.9\linewidth]{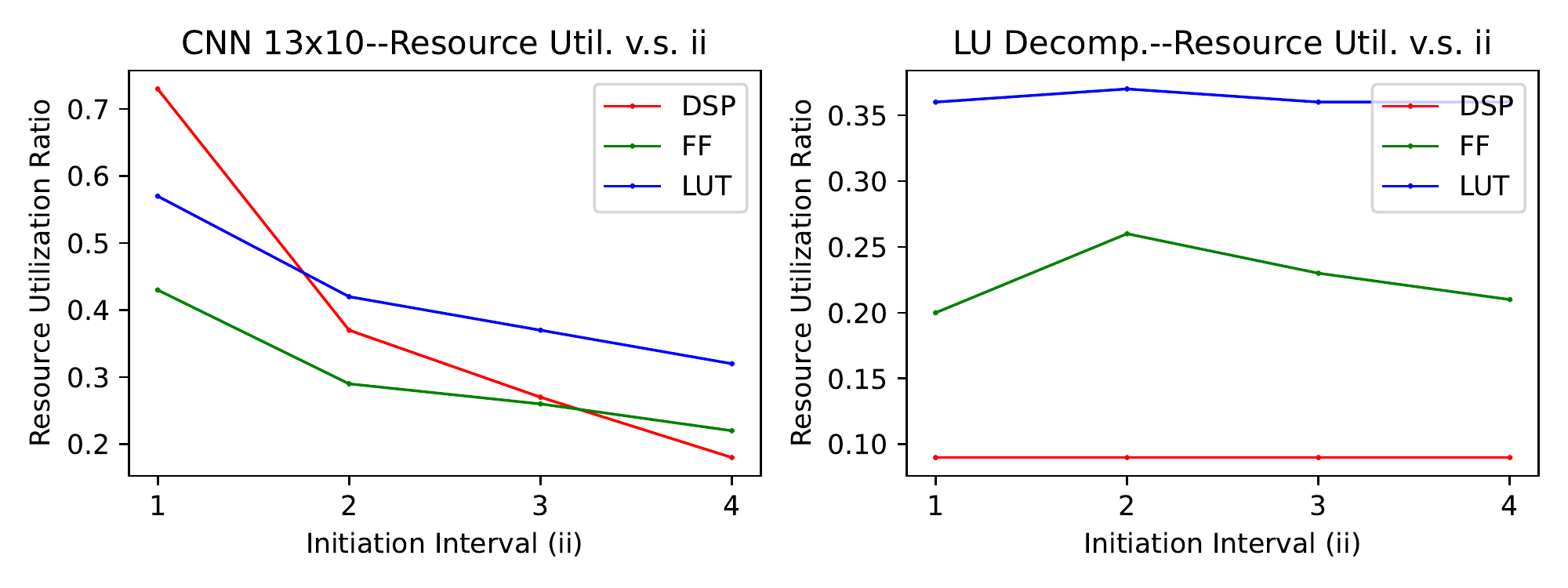}
    \vspace*{-1.5\baselineskip}
    \caption{Comparison of Resource Utilization v.s. II between CNN and LU Benchmarks.\vspace{-2\baselineskip}}
    \label{fig:5-abexample}
\end{figure}

\begin{algorithm}[!t]
\myalgsize
\caption{Top-level of FADO framework}\label{alg:5-alg1}
\DontPrintSemicolon
  \KwInput{$QoR\_lib$, $N$: look-ahead step number}
  \KwOutput{Optimal Directives, Legal Floorplan}
   \While{True } {
        $(i_1, j_1)=\argmax^1_{i,j} (L_{ij})$ \tcp*{the longest functions}
        $L_{(i_2, j_2)}=\max^2_{i, j} (L_{ij})$ \tcp*{the 2nd-longest latency}
        $DS$ = \Prune{$QoR\_lib$, $(i_{max, 1}, j_{max, 1})$, $L_{i_{max, 2}, j_{max, 2}}$};\;
        $DP$ = $DS[-1]$, break \textbf{if} $DP$ is $None$ \tcp*{Constraint Eq.~\ref{eq:4-formulation-eq3}}
        $fit$, $incrfp\_list$ = \online{$(i_1, j_1)$, DP};\;
        \If{not fit } {
            \offline, \online\;
            \If{not fit } {
                \textbf{iterative} \ahead{$N$}, \online\;
                \If{not fit } {
                    \textbf{iterative} \back{}, \online\;
                }
            }
        }
        \incrementalFloorplan{$incrfp\_list$};\;
        \exitcond;\;
   }
\end{algorithm}

\vspace*{-0.5\baselineskip}
\subsection{Directive Optimization}\label{directive-search}
\vspace*{-0.2\baselineskip}

\textbf{The Top-Level Algorithm} of floorplan-aware directive optimization is described in Alg.~\ref{alg:5-alg1}. In every iteration of floorplan-aware directive optimization, we identify the functions with the longest and second-longest latency among the whole HLS design, i.e., the sub-function $j_1$ of kernel $i_1$ is the bottleneck, represented by $(i_1, j_1)=\argmax^1_{i, j} (L_{ij})$, and the second-longest function is $(i_2, j_2)$. We apply the latency-bottleneck-guided search~\cite{liang2019hi, zhao2017comba} in \Prune{}, which extracts all design points of $(i_1, j_1)$ with smaller latency compared with the second-longest function's $L_{(i_2,j_2)}$ to form a next-step design space $DS$ (a set of directive configurations and their respective QoRs). Although applying any of the configurations in $DS$ to function $(i_1, j_1)$ would make $(i_2, j_2)$ the new latency bottleneck of the whole design, we choose the design point $DP$, which has the largest latency among $DS$, for further floorplan legalization.

To compare, FADO will not make one-off latency improvements for bottleneck functions or choose the design point with the lowest resource utilization. On the one hand, aggressive latency improvement usually results in a dramatic increase in the resource utilization of current function(s), and potential latency improvement for future bottlenecks could be precluded because of a lack of resources. On the other hand, since resource utilization is calculated by taking the maximum ratio among different resources, considering the non-monotonic design space, when utilization of one resource is minimized, others could still increase. Hence, we always assign the top priority to $DP$ for floorplanning. Note that functions having the same latency with $(i_1, j_1)$ are considered as a batch for efficiency.

\vspace*{0.5\baselineskip} 
\noindent\textbf{Look-Ahead and Look-Back.} Guided by the latency bottleneck, the main algorithm prunes the ineffective design points with negligible improvement in the overall latency. However, in realistic HLS designs, the greedy algorithm could get stuck in local optima. Fig.~\ref{fig:5-abexample} shows the different trend of resource utilization as the \textit{PIPELINE} initiation interval (II) changes in two designs, CNN from \cite{cong2018polysa} and LU from \cite{wang2021autosa}. As II increases, latency also increases in both designs. It results in less utilization of the three types of resources in the CNN benchmark because computation instances are shared among multiple cycles. However, in contrast, there are a lot of loop-carried dependencies in the LU benchmark, the results from the previous iteration cannot be directly passed to the next, and extra logic is used to buffer the results. Together with the effect of resource sharing, the utilization first increases as II increases from 1 to 2, and then decreases when II continues increasing.

To handle the non-monotonic design space, we propose the sampling methods of \ahead{} and \back{}. When the first two stages of floorplanning --- online packing and offline re-packing fail to find a legal floorplan for $DP$, we further check the floorplan for a certain number of design points with lower latency yet potentially fewer resources. This is referred to as the look-ahead stage. If it still fails in floorplanning, we turn to check the points with larger latency than $DP$ in the look-back stage. These points are more likely to have lower resource utilization and a legal floorplan.

\begin{figure}[tbp]
\vspace*{-0.1\baselineskip}
    \centering
    \includegraphics[width=0.9\linewidth]{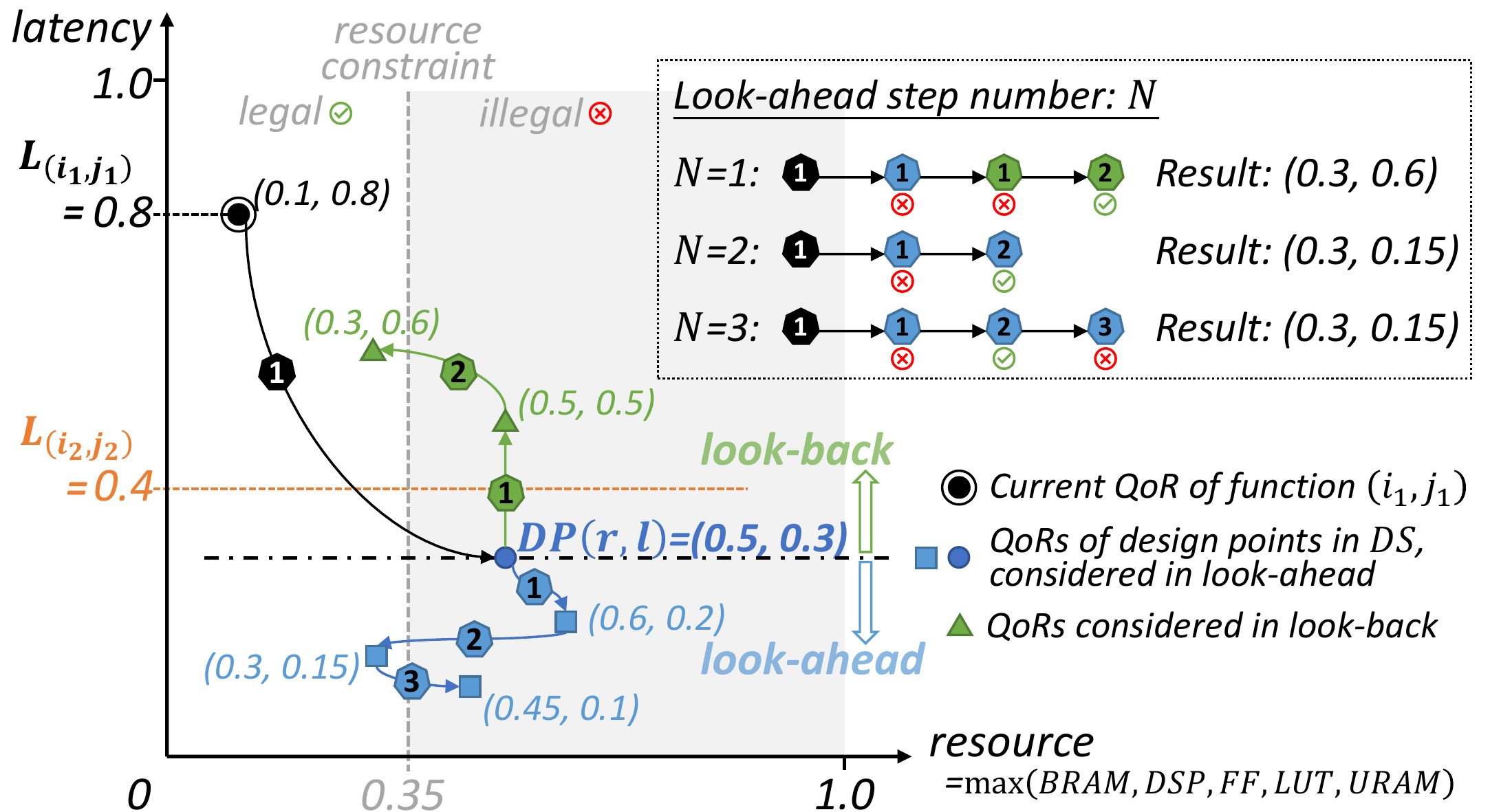}
    \vspace*{-0.8\baselineskip}
    \caption{ Look-Ahead with Step \# $N$, and Look-Back. }
    \label{fig:5-aheadback}
\end{figure}

Fig.~\ref{fig:5-aheadback} shows a snapshot of directive search for the current bottleneck function $(i_1, j_1)$. QoR values are normalized for clarity. The design point with the top priority for floorplan checking is $DP$, with a resource utilization of 0.5 and a latency of 0.3 (the longest latency smaller than $L_{(i_2, j_2)}=0.4$). However, only 0.35 resource is left for the current function, and there's no legal floorplan found for $DP$ during online packing and offline re-packing. We now look ahead/back for other improvement opportunities with fewer resources. When we set the step number $N$ to 1 during \ahead{}, the next design point consumes 0.6 utilization and also fails to be floorplanned. Thus, when \ahead{} also fails to find a point with a legal floorplan, \back{} traverses all the points with latency from 0.3 to 0.8. When $N$ is set to 2 or 3, the directive configuration with a latency of 0.15 is found during \ahead{}.

For HLS designs, our implementation decides the step number $N$ of \ahead{} by analyzing the range of parameters for \textit{PIPELINE} and \textit{UNROLL}, two of the most effective directives. We define the $N$ as the largest number of different configurations on a single nested loop. To exclude the directives over-utilizing resources, we check at most three levels for each nested loop from the innermost level. For each nested loop of $n$ levels, we index the innermost loop with $1$, and the outermost loop with $n$. (1) For directive \textit{PIPELINE}, since \cite{xilinx2020autopl} suggests a maximum loop bound of 64 for auto pipelining, we set the range of II to the logarithm of the minimum between 64 and the iteration latency $IL$ from HLS report. (2) For directive \textit{UNROLL}, similar to \textit{PIPELINE}, we take the minimum between 64 and the loop bound $B$. (3) For the combination of \textit{PIPELINE} and \textit{UNROLL}, since all the inner loops are completely unrolled when an outer loop is pipelined, we consider only the directive combination in 2 levels of loops. In all, the resulting step number $N$ for a design with $m$ nested loops is:
\newpage

\vspace*{-2\baselineskip}
\begin{equation}\label{eq:5-ahead}
  \begin{aligned}
    & N_1&&=\max_{1\leq i\leq m} \sum_{j=1}^{\max(3, n_i)}\log_2\min(64, {IL}_{ij})\\[-5pt]
    & N_2&&=\max_{1\leq i\leq m} \sum_{j=1}^{\max(3, n_i)}\log_2\min(64, B_{ij})\\[-5pt]
    & N_3&&=\max_{1\leq i\leq m} \sum_{j=1}^{\max(2, n_i)}\log_2\min(64, B_{ij})\\[-5pt]
    & N  &&=N_1+N_2+N_3\\
  \end{aligned}
\end{equation}
\vspace*{-1\baselineskip}

\begin{algorithm}[!t]
\myalgsize
\caption{ Online Packing }\label{alg:5-alg2}
\DontPrintSemicolon
  \KwInput{$longest\_functions$ and QoRs $L_{ij}, R_{ij}$, design point $p$}
  \KwOutput{$fit$, $incrfsp\_list$}
    $fit$ = False, $incrfp\_list$ = [], $unfit\_funcs$ = [];\;
    \For{$func$ in longest\_functions }{
        \If{$func$ still fits in the current slot $s_c$ }{
            update directives for $func$, and resource util. for $s_c$;\;
            $L_{ij}, R_{ij} = L_{p}, R_{p}$; $fit$ = True;\;
        }
        \Else{
            \tcp{check constraint Eq.~\ref{eq:4-formulation-eq6}, Eq.~\ref{eq:4-formulation-eq7} and Eq.~\ref{eq:4-formulation-eq8}}
            calculate overflow ratio, and sort $other\_slots$ by $CR$;\;
            \For{each $s_o$ in $other\_slots$ }{
                \If{no overflow when moving $func$ to $s_o$ }{ 
                    \tcp{check constraint Eq.~\ref{eq:4-formulation-eq5}}
                    update directives for $func$;\;
                    update util. for $s_o$, $s_c$; \tcp*{constraint Eq.~\ref{eq:4-formulation-eq4}}
                    append ($func$, $s_o$) to $incrfp\_list$;\;
                    $L_{ij}, R_{ij} = L_{p}, R_{p}$; $fit$ = True; break;\;
                }
            }
            \If {not fit } {
                append $func$ to $unfit\_funcs$;\;
            }
        }
    }
    \If{any func in unfit\_funcs }{
        clear $incrfp\_list$; $fit$ = False;\;
    }
\end{algorithm}

\vspace*{-0.5\baselineskip}
\subsection{Incremental Floorplanning}\label{incr-fp}
\vspace*{-0.2\baselineskip}

The initial floorplan input to FADO is generated by an iterative min-cut MILP bi-partitioning in the "AutoBridge Floorplanner~\cite{guo2021autobridge}" shown in Fig.~\ref{fig:1-framework}. During FADO's iterations, we apply a resource-bottleneck-guided online WF algorithm. When the online packing fails in finding a legal floorplan, an offline BFD re-packing compacts the existing floorplan before calling the online packing again. The definition of "online" and "offline" algorithms refers to \cite{karp1992line}. During online packing, HLS functions are optimized one after another, and the previous floorplan of a function will be kept unchanged. In contrast, without applying new directives, the offline stage reorder all functions by heuristics to improve the packing quality.

\vspace*{-0.5\baselineskip}
\subsubsection{Online Packing}\label{online-fp}

To avoid routing congestion from a high-abstraction view, the online packing tends to balance the utilization ratio among different resources and different slots, i.e., if a function fails to fit into its original slot after applying a new directive configuration, we try to floorplan it into other slots according to the non-decreasing order of critical resource (CR). CR refers to the type of resource having the most overflow percentage among BRAM, DSP, FF, LUT, and URAM. If there are multiple slots with the same CR, we sort them by the average utilization of the other four non-critical resources. The online packing algorithm is shown in Alg.~\ref{alg:5-alg2}. Note that overflow ratios are calculated by each resource.

\begin{figure}[tbp]
    \centering
    \vspace*{-0.8\baselineskip}
    \includegraphics[width=0.88\linewidth]{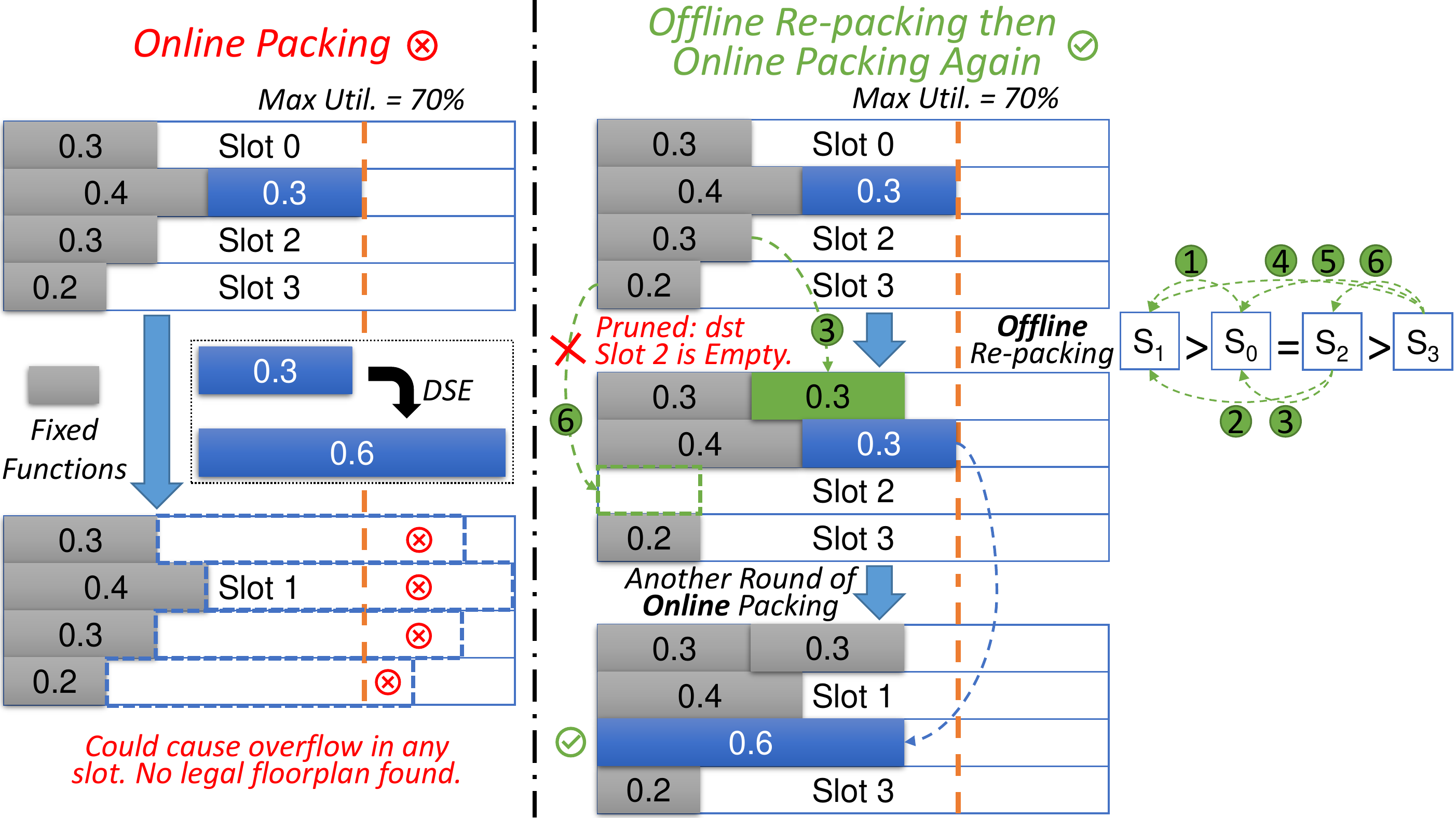}
    \vspace*{-0.8\baselineskip}
    \caption{ An Example of Offline Re-packing. }
    \label{fig:5-offline}
\end{figure}

\begin{figure}[tbp]
\vspace{-1\baselineskip}
    \centering
    \includegraphics[width=0.63\linewidth]{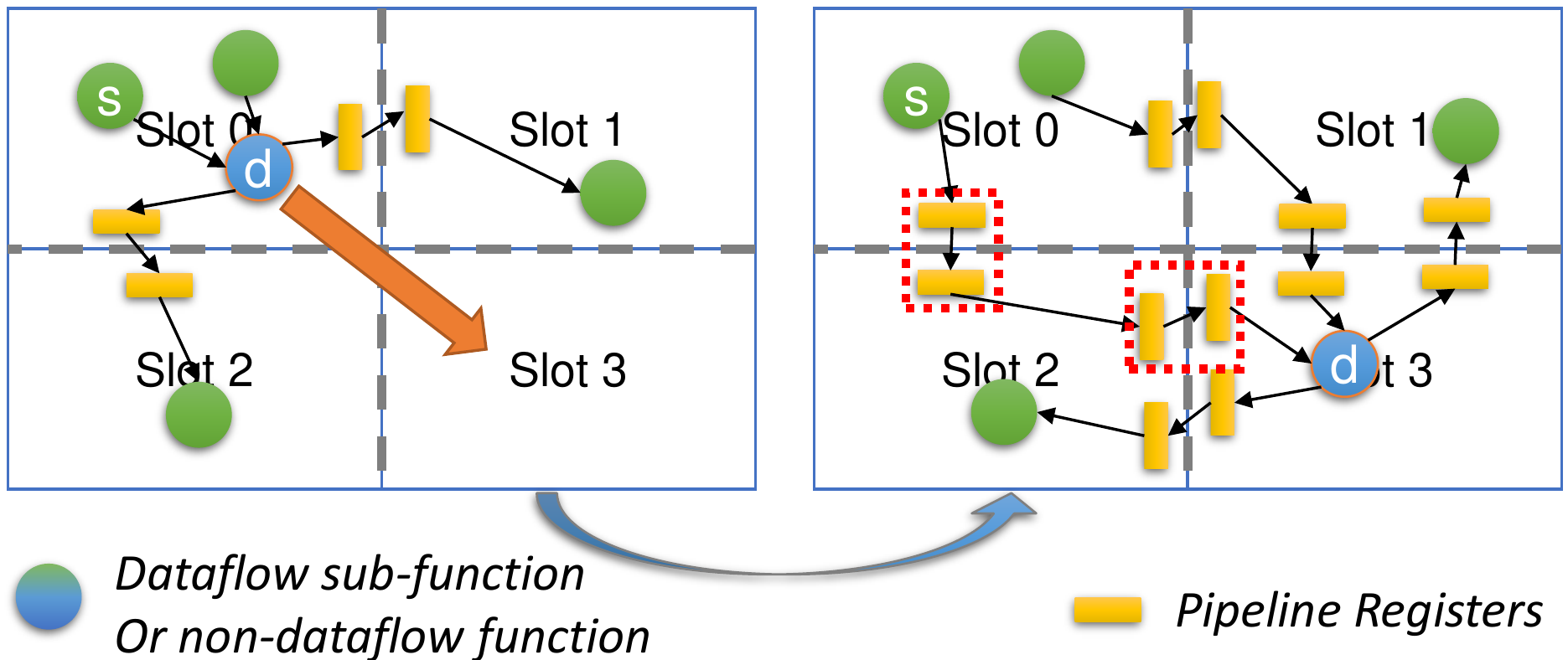}
    \vspace*{-1\baselineskip}
    \caption{ Incremental Update of Pipeline Registers. }
    \label{fig:5-incrpl}
\end{figure}

\vspace*{-0.5\baselineskip}
\subsubsection{Offline Re-packing}\label{offline-fp}

Since \online{} tends to spread functions evenly on each slot, and when there's an aggressive move in directive search with a sharp increase in resource utilization, the balance could preclude the new design point from taking effect. Thus, offline re-packing sorts all the slots ${SL}_i$ by resource utilization in non-increasing order. Then, it respectively sorts all the functions $F_{ij}$ on each slot ${SL}_i$ by resource in non-increasing order as well. The re-packing starts with moving the $F_{21}$ from the second fullest slot ${SL}_2$ to the fullest ${SL}_1$, and then the second largest function $F_{22}$ from ${SL}_2$ to ${SL}_1$, etc. When ${SL}_1$ is full, or ${SL}_2$ is empty, we turn to move functions $F_{31}, F_{32},$ etc., from the third fullest slot ${SL}_3$ to ${SL}_1$, then to ${SL}_2$. A general step of re-packing the $m$-th fullest slot is to move $F_{m1}, F_{m2}, ..., F_{mn_m}$ to ${SL}_1, {SL}_2, ..., {SL}_{m-1}$ in turn.

Fig.~\ref{fig:5-offline} shows the floorplanning of 5 functions onto 4 slots. We reduce multiple resources to one dimension by taking the maximum ratio among them. Through directive search, the resource of the blue function expands from 0.3 to 0.6. However, since other gray functions were fixed during the online stage, the expanded function fits nowhere. To compare, the re-packing stage sorts the slots by utilization in non-increasing order and executes six trials in order. Trials 1/2 fail because slot 1 is full. During trial 3, a gray function is moved from slot 2 to slot 0. Trials 4/5 also fail because the destinations, slots 0 and 1, are full. Trial 6 is canceled because the destination slot 2 has been found empty. With re-packing, the expanded function fits in slot 2 after a new round of online packing.

Re-packing applies different strategies for functions in dataflow kernels and non-dataflow kernels. The floorplan change is free for dataflow functions if the SLL utilization constraint is met. For non-dataflow parts, since their connections with other adjacent functions are not through the FIFO channel, the long wires could not be broken by inserting pipeline logic. Hence, instead of moving them, we try to force other dataflow functions that have no RAM connection with them to different slots. Thus, FADO assigns more resources to the slots containing non-dataflow for further DSE.

\vspace*{-0.8\baselineskip}
\subsection{Incremental Pipelining}\label{incr-pl}

When moving a function across slots, as Fig.~\ref{fig:5-incrpl} shows, each time a path crosses a boundary between two slots (SLR boundary or I/O banks), additional pipeline registers should be added beside the boundaries to break down the long wires. We here set a constraint of 90\% (Eq.~\ref{eq:4-formulation-eq8}) for SLL utilization and incrementally update the pipelining logic of long wires crossing slot boundaries. As Fig.~\ref{fig:5-incrpl} shows, when function "d" is moved from Slot 0 to Slot 3, two groups of additional pipeline registers are added between "s" and "d".

\vspace*{-0.6\baselineskip}
\subsection{Exit Condition and External Tools}\label{external}
\vspace{-0.1\baselineskip}
During iterative optimization, when there's no legal floorplan found for the next design point of the current longest function or when no other directive configuration could improve the bottleneck's latency further, it is excluded in future iterations. FADO stops when there's no function left for the bottleneck analysis. Then, it dumps the optimal directives to a TCL file to guide the re-synthesis of the HLS code to generate a high-performance RTL design. FADO also delivers the final floorplan to the global router, latency balancer, and dataflow RTL generator within \cite{guo2021autobridge} to update the pipelining of dataflow kernels in the Verilog code, and generate another TCL script to guide the floorplanning during implementation in Vitis.

\vspace{-0.5\baselineskip}

%% file: sec-results.tex
\section{Results}\label{sec-results}

\begin{table*}[tbp]
\vspace*{-0.6\baselineskip}
    \centering
    \setlength\tabcolsep{2pt}
    \caption{ QoR Comparison between FADO and Other DSE Strategies \vspace{-1\baselineskip}}{
    \resizebox{\textwidth}{!}{
    \begin{tabular}{c || c c c c | c || c c c c |c || c c c c |c} 
        \toprule
        Benchmarks & \multicolumn{5}{c||}{CNN*2+2MM*1} & \multicolumn{5}{c||}{MM*1+COV*2} & \multicolumn{5}{c}{MTTKRP*2+HEAT*2} \\
        \midrule
        & Resource$^a$ & Runtime$^b$ & Latency$^c$ & Fmax$^d$ & \textbf{Exe\_time}$^g$ & Resource & Runtime & Latency & Fmax & \textbf{Exe\_time} & Resource & Runtime & Latency & Fmax & \textbf{Exe\_time} \\
        \hline
        Original (no directive) & 28\% & - & 8933 & - & - & 20\% & - & 131839 & - & - & 57\% & - & 8147919 & - & - \\
        Initial FP$^e$ -> Iterative DO$^f$ & 28\% & 2.24 & 735 & \textbf{300.45} & 2,445 & 19\% & 0.16 & 131839 & \textbf{282.86} & 466,094 & 46\% & 1.36 & 8138605 & \textit{Failure} & - \\ 
        Iterative (DO + AutoBridge FP) & 48\% & 1658 & 92.7 & 235.89 & 393 & 41\% & 10307 & 2549 & 278.84 & 9,141 & 62\% & 3554 & 598532 & 159.97 & 3,741,525 \\
        Iterative (DO + Incr FP) (\textbf{Ours}) & 55\% & 2.39 & \textbf{91.2} & 269.95 & \textbf{338} & 41\% & 2.17 & \textbf{1647} & 274.47 & \textbf{6,001} & 63\% & 1.95 & \textbf{128104} & \textbf{300.45} & \textbf{426,374} \\
        \bottomrule
        \toprule
        Benchmarks & \multicolumn{5}{c||}{MM*2+2MM*2} & \multicolumn{5}{c||}{CNN*3+COV*2} & \multicolumn{5}{c}{MTTKRP*2+COV*2} \\
        \midrule
        & Resource & Runtime & Latency & Fmax & \textbf{Exe\_time} & Resource & Runtime & Latency & Fmax & \textbf{Exe\_time} & Resource & Runtime & Latency & Fmax & \textbf{Exe\_time} \\
        \hline
        Original (no directive) & 40\% & - & 259516 & - & - & 31\% & - & 18130 & - & - & 38\% & - & 8113234 & - & - \\     
        Initial FP -> Iterative DO & 59\% & 1.13 & 258842 & 274.10 & 944,335 & 39\% & 2.03 & 6716 & \textbf{300.45} & 22,354 & 42\% & 2.18 & 8113234 & \textbf{300.45} & 27,003,607 \\
        Iterative (DO + AutoBridge FP) & 60\% & 32656 & 67652 & \textit{Failure} & - & 62\% & 8301 & 1278 & 222.01 & 5,754 & 61\% & 12627 & 562017 & \textbf{300.45} & 1,870,585 \\
        Iterative (DO + Incr FP) (\textbf{Ours}) & 58\% & 6.63 & \textbf{66158} & \textbf{300.00} & \textbf{220,527} & 63\% & 5.04 & \textbf{1233} & \textbf{300.45} & \textbf{4,105} & 64\% & 4.89 & \textbf{126921} & \textbf{300.45} & \textbf{422,437} \\
        \bottomrule
        \multicolumn{16}{l}{$^a$ The maximum utilization ratio among BRAM, DSP, FF, LUT, and URAM. $^b$ DSE runtime in seconds. $^c$ Execution time of HLS designs in number of \emph{thousand} clock cycles. }\\ 
        \multicolumn{16}{l}{$^d$ Maximum achievable frequency in MHz.  $^e$ FP: Floorplanning. $^f$ DO: Directive Optimization. $^g$ Overall execution time (cycle number/frequency) of HLS designs in microseconds ($\mu$s). }\\
    \end{tabular}}
    }
    \vspace*{-1.2\baselineskip}
    \label{tab:6-compare-result}
\end{table*}

\begin{figure}[tbp]
\vspace{-0.8\baselineskip}
    \centering
    \includegraphics[width=0.82\linewidth]{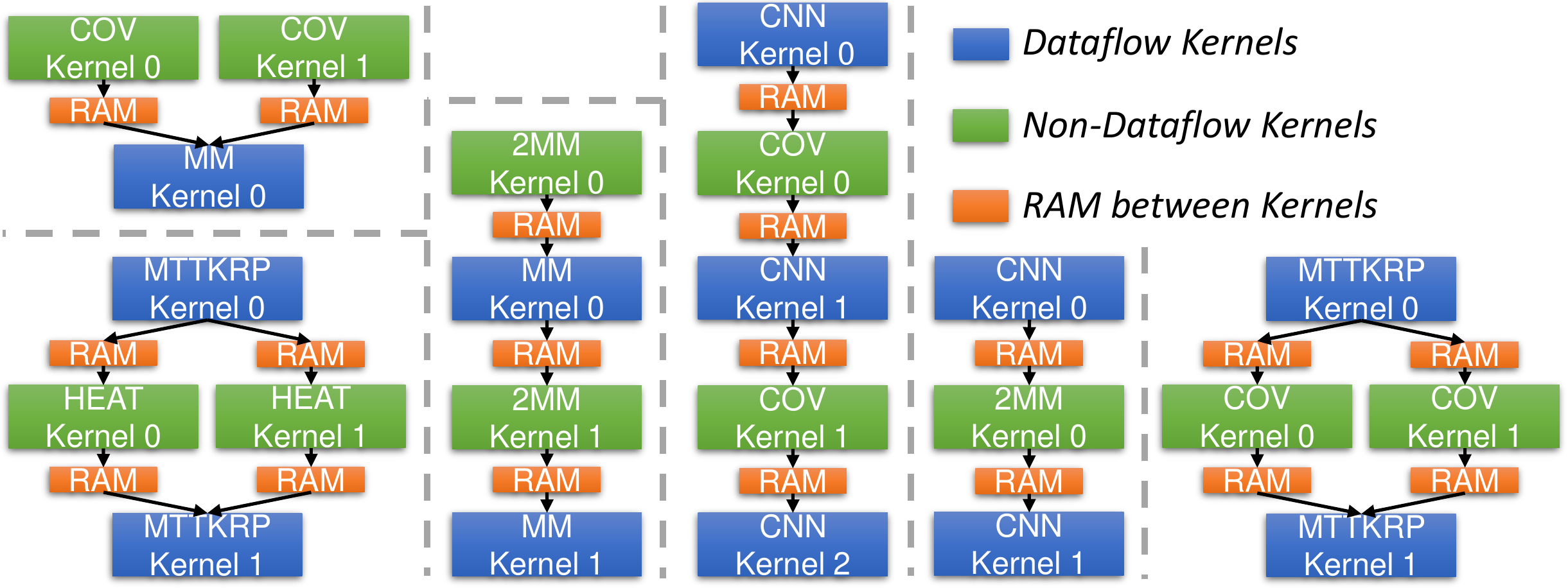}
    \vspace*{-0.8\baselineskip}
    \caption{ The 6 Benchmarks Used to Evaluate FADO. }
    \label{fig:6-benchmarks}
\end{figure}

\vspace{-0.2\baselineskip}
\subsection{Benchmarks and Experiment Settings}\label{sec-6-benchmark}

We mainly adopt large-scale open-source HLS designs with compatible interfaces for evaluation and filter out many commonly used but unsuitable benchmarks. To be specific, most of the designs in Vitis Libraries~\cite{vitislib}, CHstone~\cite{hara2008chstone}, Rosetta~\cite{rosetta}, etc., occupy less than 10\% of resources on the Alveo U250 FPGA. They only have several functions to consider during coarse-grained floorplanning, which is not challenging even if we increase the design size, e.g., by applying a larger bitwidth. Besides, interface incompatibility makes it difficult to scale up by connecting multiple designs from these benchmarks. Hence, we generate large dataflow \textit{CNN}, \textit{MM} and \textit{MTTKRP} using PolySA~\cite{cong2018polysa} and AutoSA~\cite{wang2021autosa}. As for non-dataflow designs, we use \textit{2MM}, \textit{COV}, and \textit{HEAT} from Polybench~\cite{polybench}, which are general programs also used in CPU, GPU, etc. To best show the generality of our solution, we assemble six large benchmarks mixing the dataflow and non-dataflow kernels above to evaluate the performance of our framework, as Fig.~\ref{fig:6-benchmarks} shows. The kernels connect through RAMs, which enlarges the design space compared with a single dataflow kernel.

\begin{figure}[tbp]
    \centering
    \includegraphics[width=0.78\linewidth]{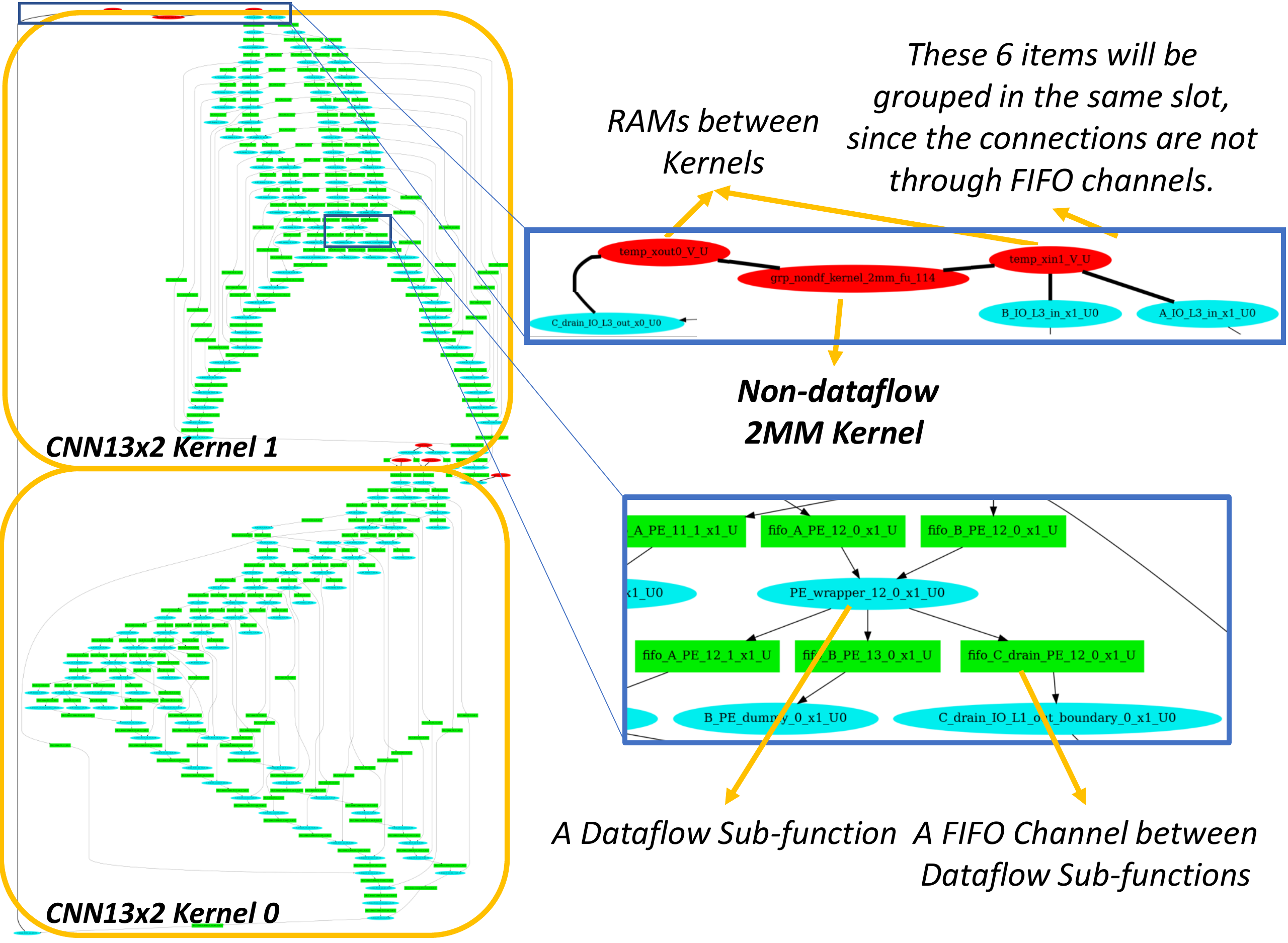}
    \vspace*{-0.7\baselineskip}
    \caption{ The Scale of the CNN*2+2MM*1 Benchmark.  }
    \label{fig:cnn2mm_scale}
\end{figure}

To show the scale of our problem, we visualize the HLS-function-level data flow graph of the \textit{CNN*2+2MM*1} benchmark in Fig.~\ref{fig:cnn2mm_scale}. The two yellow bounding boxes mark the two \textit{CNN13x2} dataflow kernels, each containing hundreds of sub-functions. The red circles on the top of this figure are the non-dataflow \textit{2MM} kernel and the two RAMs connected to it. The RAM module "temp\_xin1\_V\_U" is connected to two input sub-functions of \textit{CNN13x2} Kernel 1, and RAM "temp\_xout0\_V\_U" is connected to one output sub-function of \textit{CNN13x2} Kernel 0. Since the connection among them are not through FIFO channels, they are grouped during floorplanning and always placed in the same slot. As for a dataflow kernel, the green boxes are FIFO channels, and the blue circles are dataflow sub-functions. Dataflows can be partitioned, floorplanned, and pipelined on any slot as long as the resource constraints are met. The overall design space of FADO is the Cartesian product of directive space and floorplan space. For directive search, the space ranges from millions to billions in our benchmarks, considering the parameters in Table 4. For floorplanning, it maps hundreds of functions to four slots, and the space size is four to the power of hundreds.

We use the Xilinx Vitis HLS 2020.2 for HLS synthesis and Vitis for implementation. We evaluate our framework on the Xilinx Alveo U250 FPGA, which contains eight slots defined by the 4 SLRs and an I/O bank in the middle. Note that the rightmost column of clock regions is occupied by Vitis platform IP. Hence the resource calculation excludes that column. We reduce the floorplanning of HLS designs to the lower half\footnote{The lower half of Alveo U250 FPGA excluding the rightmost column of clock region contains 2016 BRAMs, 5184 DSPs, 1319040 FFs, 659520 LUTs, and 544 URAMs.} (4 slots on SLR 0 and SLR 1) of the FPGA to enable exhaustive floorplan search in analyzing the optimality of our results. In our experiments, we find the resource constraint of 70\% still leads to placement or routing failure sometimes. Hence, we tighten the limit to 65\% for each slot during DSE.

\vspace{-0.5\baselineskip}
\subsection{Comparative Experiments}\label{sec-6-compare}

Table.~\ref{tab:6-compare-result} compares FADO with different top-level DSE algorithms and floorplanning algorithms. We report both the DSE runtime and the quality of each design implementation with their latency, maximum frequency, and overall execution time. Among these metrics, the overall execution time combines latency and timing quality, reflecting the ultimate design performance on FPGA.

In Table.~\ref{tab:6-compare-result}, the "Initial FP -> Iterative DO" baseline performs the directive optimization using one-off initial floorplanning. It applies the min-cut MILP floorplanning from \cite{guo2021autobridge}, and all HLS functions' positions are fixed during the iterative directive search. The limited optimization opportunities caused by the fixed initial floorplan lead to an under-utilization of resources. This seriously limits the latency optimization, resulting in the longest latency for all benchmarks. It fails in the implementation of \textit{MTTKRP*2+HEAT*2} because two \textit{HEAT} kernels are floorplanned on the same slot, but each has a large array using more than one column of BRAM or URAM, which triggers an exception during placement. 

\begin{table*}[tbp]
\vspace{-0.7\baselineskip}
    \centering
    \caption{ Stages of Floorplan-aware Directive Optimization \vspace{-1\baselineskip}}{
    \resizebox{0.78\textwidth}{!}{
    \begin{tabular}{c || c c || c c || c c || c c || c c || c c} 
        \toprule
        Benchmarks & \multicolumn{2}{c||}{CNN*2+2MM*1} & \multicolumn{2}{c||}{MM*1+COV*2} & \multicolumn{2}{c||}{MTTKRP*2+HEAT*2} & \multicolumn{2}{c||}{MM*2+2MM*2} & \multicolumn{2}{c||}{CNN*3+COV*2} & \multicolumn{2}{c}{MTTKRP*2+COV*2}\\
        \midrule
        Stages & Resource$^a$ & Latency$^b$ & Resource & Latency & Resource & Latency & Resource & Latency & Resource & Latency & Resource & Latency \\
        \hline
        Online & 28.27\% & 735 & 40.66\% & 5167 & 63.15\% & 163241 & 40.28\% & 259516 & 31.79\% & 4718 & 62.95\% & 141260 \\
        Offline & 40.12\% & 132 & 40.66\% & 5167 & 64.67\% & 153927 & 40.28\% & 259516 & 31.79\% & 4718 & 62.95\% & 141260 \\
        Look-Ahead & 55.01\% & 91.4 & 40.66\% & 1651 & 63.26\% & 129184 & 55.25\% & 66184 & 31.79\% & 4718 & 64.49\% & 126921 \\
        Look-Back & 54.56\% & 91.2 & 40.66\% & 1647 & 63.25\% & 128104 & 57.53\% & 66158 & 63.32\% & 1233 & 64.49\% & 126921 \\
        \bottomrule
        \multicolumn{13}{l}{$^a$ The maximum utilization ratio among BRAM, DSP, FF, LUT, and URAM. $^b$ Execution time of HLS designs in number of \emph{thousand} clock cycles.}\\
    \end{tabular}}
    }
    \label{tab:6-stages}
    \vspace{-1\baselineskip}
\end{table*}

\begin{figure*}[tbp]
    \centering
    \includegraphics[width=0.8\textwidth]{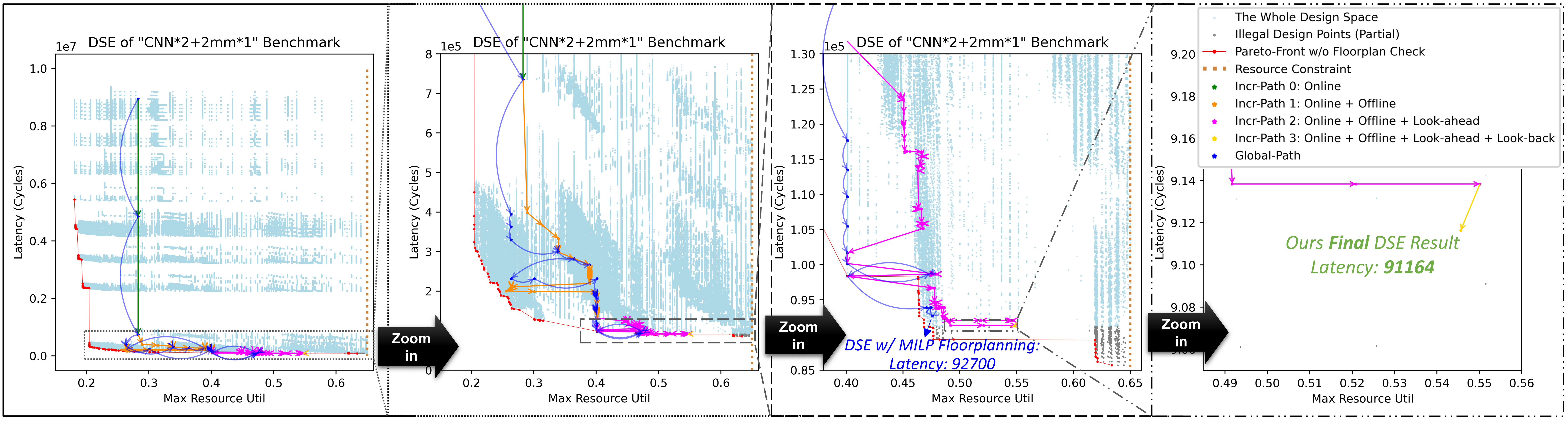}
    \vspace{-1.5\baselineskip}
    \caption{ DSE Stages and Results on the CNN*2+2MM*1 Benchmark. \vspace{-1.4\baselineskip} }
    \label{fig:6-cnn-result}
\end{figure*}

The "Iterative (DO + AutoBridge FP)" baseline runs the min-cut MILP floorplanning iteratively. Note that the heuristics of look-ahead and look-back are also applied in this algorithm for fairness when compared with FADO. This algorithm results in orders-of-magnitude longer runtime than FADO due to repetitively calling the MILP solver. Meanwhile, since AutoBridge~\cite{guo2021autobridge} applies iterative bi-partitioning rather than one-off eight-way partitioning\footnote{Eight-way partitioning runs even more than 10x slower compared with bi-partitioning in directive-floorplan co-search experiments using benchmarks above.}, it fails to reach some of the solutions. As reflected by the execution time, its design implementation quality is inferior to FADO in all six benchmarks. In summary, this method takes a significantly longer time while still resulting in a sub-optimal design.

As for our FADO, the online packing and offline re-packing strategies alternatively balance and compact the floorplan, contributing to full utilization of resources on multiple dies (the highest utilization ratio under resource constraint of 65\% in five out of all six benchmarks). Accordingly, the high-quality floorplan provides strong support for exploring a larger design space during the directive search, thus our FADO achieves 33.12\% smaller latency on average compared with the time-consuming "iterative (DO + AutoBridge FP)", and attains the lowest latency for all benchmarks over all baselines. The latency improvement varies because of the nature of benchmarks -- it's more significant when FADO legalizes the floorplan for some bottleneck functions with a great latency-resource tradeoff, as the cases \textit{MM*1+COV*2}, \textit{MTTKRP*2+HEAT*2}, and \textit{MTTKRP*2+COV*2} show. As for frequency, experiments show that when the utilization gets close to 65\%, although the frequency could vary to some extent due to non-determinism in floorplanning and further implementation, our incremental solution still outperforms the baselines, with both a higher average Fmax of 290.96 MHz and lower variance. Moreover, since our incremental legalization leads to a minimum change of floorplan in each iteration of co-optimization, it's much more efficient than updating all functions' locations globally. This efficient legalization algorithm contributes to a speedup of 693X$\sim$4925X in the runtime of the entire co-optimization. Without any loss in timing quality, the design implementation quality reflected in overall execution time is 1.16X$\sim$8.78X better than the best baseline.

\vspace{-0.8\baselineskip}
\subsection{Analysis of DSE Stages}\label{sec-6-stages}
\vspace{-0.3\baselineskip}

Fig.~\ref{fig:6-cnn-result} shows the multiple stages of directive-floorplan co-search for \textit{CNN*2+2MM*1} benchmark. The horizontal axis takes the maximum utilization among resources on the FPGA, and the vertical axis shows the latency in the number of clock cycles. The cyan points represent the whole directive design space without floorplan legality check, with red dots showing the Pareto-front. Our search starts from the highest latency point (28.27\%, 8933000). In the first stage, the green arrows at the beginning are showing online floorplanning. It stops at (28.27\%, 734592) because of a sharp resource increase of the large non-dataflow kernel. In the second stage, the offline re-packing clears out the dataflow sub-functions on the least-occupied slot, and continues until (40.12\%, 131752), where the yellow arrows fall into a local maximum resource in the second sub-figure. It would fail to continue without looking ahead for points with less utilization of the current critical resource. As the third sub-figure shows, the pink arrows reach (55.01\%, 91384) by looking ahead for three more design points after failure in previous stages. Finally, the DSE stops at (54.56\%, 91164) after the final look-back. To compare, the DSE with global MILP floorplanning stops earlier at (47.59\%, 92700). To show the optimality of our result, we check the floorplan legality for all design points with less latency than our result of 91164 --- all the gray points have no legal floorplan when running global MILP floorplanning solely.

Table.~\ref{tab:6-stages} shows the DSE results of different optimization stages in FADO. Note that the four stages are running sequentially in each iteration, and the latency/resource in this table is not the result of each stage acting alone, except for "Online". For example, for stage "Look-Ahead", it includes the joint effort of (1) online packing, (2) offline re-packing followed by another round of online packing, and (3) look-ahead followed by online packing, as described in Alg.~\ref{alg:5-alg1}. It's possible that for some iterations, we only use (1), or (1)+(2), while using (1)+(2)+(3) in the worst cases. The QoR of each stage shown in Table 6 measures the legal design point with the smallest latency achieved before the first call to the next stage. For example, the results for "Look-Ahead" is the legal point with the smallest latency achieved before the first call to \back{}.


For benchmarks \textit{CNN*2+2MM*1} and \textit{MTTKRP*2+HEAT*2}, each stage is more effective than the previous ones to avoid local optima. However, the offline method fails to improve the results in \textit{MM*1+COV*2}, \textit{MM*2+2MM*2}, and \textit{MTTKRP*2+COV*2}, compared with online stage. This happens when there are large design points with over-utilization. For example, the non-dataflow \textit{COV} kernel consumes 30 DSPs when without any directive. However, when we unroll the loop containing the multiplication operation, the DSP increases to 1920, which is more than the total DSP available in any slot. Thus, offline stage fails to optimize the floorplan, and the bottleneck, DSP utilization always remains the same value during DSE in \textit{MM*1+COV*2} and \textit{MTTKRP*1+COV*2}. For \textit{CNN*3+COV*2}, since \textit{COV} kernel has a longer latency than \textit{CNN}, major improvements are enabled by the look-back applied at the beginning of DSE.

\vspace{-0.85\baselineskip}
\subsection{Optimality Analysis}\label{sec-6-optimal}
\vspace{-0.3\baselineskip}

To analyze the optimality of our latency achieved, we check the floorplan legality for all design points with less latency than our final result. For those benchmarks with too many design points to check legality, we sample 2000 points from them and run MILP floorplanning respectively for each point. Our results show that there's no legal floorplan when a design point's latency is below our final result for all six benchmarks.

\vspace{-0.85\baselineskip}

%% file: sec-conclusion.tex
\section{Conclusion}\label{sec-conclusion}
\vspace{-0.3\baselineskip}

Our work produces FADO, an open-source framework that co-optimizes the directives and floorplan of HLS designs implemented on multi-die FPGAs. FADO combines a latency-bottleneck-guided directive optimization and an incremental floorplanning algorithm mixing various bin-packing heuristics. On the one hand, our well-customized incremental floorplanning achieves a speedup of 693X $\sim$4925X over the global MILP floorplanning~\cite{guo2021autobridge}. On the other hand, our co-optimization enables full utilization of resources on multiple dies and greatly benefits both the latency and timing. Among all six large-scale benchmarks mixing dataflow and non-dataflow kernels, FADO optimizes their execution time with a speedup of 1.16X$\sim$8.78X compared with the global floorplanning solution.

\vspace{-0.8\baselineskip}